\documentclass[a4paper,11pt]{article}
\pdfoutput=1 

\usepackage{jheppub} 
\usepackage[T1]{fontenc} 
\usepackage{slashed}
\usepackage{subcaption}
\usepackage{ucs}
\usepackage{hyperref}
\usepackage[utf8x]{inputenc}
\usepackage{amsmath,bm}
\usepackage{amsfonts}
\usepackage{comment}
\usepackage{amssymb}
\usepackage{array}
\usepackage{epsfig}
\usepackage{slashed}
\usepackage{lipsum}
\usepackage{hyperref}
\hypersetup{
	colorlinks=true,
	linkcolor=blue,
	filecolor=red,
	urlcolor=blue,
	citecolor=red
} 

\title{\boldmath Momentum space parity-odd CFT 3-point functions}

\author[a]{Sachin Jain,}
\author[a]{ Renjan Rajan John,}
\author[a]{Abhishek Mehta,}
\author[c]{Amin A.~Nizami,}
\author[a]{Adithya Suresh}

\affiliation[a]{Indian Institute of Science Education and Research, Homi Bhabha Road, Pashan, Pune 411 008, India}
\affiliation[c]{Department of Physics, Ashoka University, India}

\emailAdd{sachin.jain@iiserpune.ac.in}
\emailAdd{renjan.john@acads.iiserpune.ac.in}
\emailAdd{aan27cam@gmail.com}
\emailAdd{\{abhishek.mehta,s.adithya\}@students.iiserpune.ac.in}
\abstract{We study the parity-odd sector of 3-point functions comprising of scalar operators and conserved currents in conformal field theories in momentum space. We use momentum space conformal Ward identities as well as spin-raising and weight-shifting operators to fix the form of these correlators. We discuss in detail the regularisation of divergences and their renormalisation using specific counter-terms.}

\begin{document}
\maketitle
\raggedbottom
\flushbottom

\section{Introduction}
Conformal Field Theories (CFTs) have wide applicability in diverse areas of physics, and are central to our understanding of quantum field theory in terms of RG flows.
While CFTs are well studied in position space and Mellin space, they are relatively less studied in momentum space. Recent works on aspects of momentum space CFTs include \cite{Coriano:2013jba,Bzowski:2013sza,Bonora:2015nqa,Bonora:2015odi,Bonora:2016ida,sissathesis,Bzowski:2015pba,Bzowski:2015yxv,Bzowski:2017poo,Coriano:2018bbe,Bzowski:2018fql,Gillioz:2018mto,Coriano:2018tgn,Albayrak:2018tam,Farrow:2018yni,Isono:2018rrb, Isono:2019wex,Isono:2019ihz,Maglio:2019grh,Gillioz:2019lgs,Bzowski:2019kwd,gillioz2019convergent,Bautista:2019qxj,Albayrak:2019yve,Coriano:2019nkw,Albayrak:2019asr,Lipstein:2019mpu,Gillioz:2020mdd,Gillioz:2020wgw,Albayrak:2020isk,Bzowski:2020kfw,Jain:2020rmw,Jain:2020puw,Coriano:2020ccb,Albayrak:2020bso,Albayrak:2020fyp,Serino:2020pyu}. CFTs in momentum space find applications in cosmology \cite{Mata:2012bx,Ghosh:2014kba,Kundu:2014gxa,Arkani-Hamed:2015bza,Maldacena:2011nz, Arkani-Hamed:2018kmz,Sleight:2019mgd,Sleight:2019hfp,Baumann:2019oyu,Baumann:2020dch}, condensed matter physics \cite{Huh:2013vga, Chowdhury:2012km}, study of anomalies \cite{Gillioz:2016jnn,Coriano:2017mux,Gillioz:2018kwh, Coriano:2018zdo,Coriano:2020ees}, Hamiltonian truncation methods for strongly coupled field theories \cite{Katz:2016hxp, Anand:2019lkt} and, of course, the conformal bootstrap program \cite{Polyakov:1974gs, gillioz2019convergent,Isono:2019ihz,Isono:2019wex}. 

From the perspective of perturbative field theory which is naturally formulated in momentum space, it is of interest to study CFTs in the same setting. Flat space scattering amplitudes are, via AdS/CFT, directly related to the flat space limit of CFT correlators in momentum space \cite{Raju:2012zr} \footnote{There are analogous, though somewhat less straightforward relations in Mellin space \cite{Penedones:2010ue, Fitzpatrick:2011hu} and position space \cite{Gary:2009ae, Gary:2009mi, Komatsu:2020sag}.}.  Studying momentum space CFT correlators can therefore shed light on the structure of flat space amplitudes. 
Interestingly, evidence for the double copy structure - which exists for flat space amplitudes - was seen directly in momentum space CFT 3-point correlators in \cite{Farrow:2018yni, Lipstein:2019mpu}. An important simplification in momentum space is that  4-point conformal blocks can be constructed from 3-point functions in a relatively  straightforward manner in momentum space \cite{gillioz2019convergent, Gillioz:2020wgw}. 
Another significant application is in the cosmological setting where the CMB bispectrum which is a measure of non-gaussianity is given by the 3-point function in momentum space.

Three-point functions of scalar and spinning operators have a simple, well-known form in position space - this is most easily seen by going to the embedding space.
However, their momentum space analogues are quite complicated. For example, the scalar 3-point correlator in momentum space is solved in terms of triple-$K$ integrals which can involve divergences \cite{Bzowski:2013sza}. A careful treatment would require the regularisation and renormalisation of these divergences \cite{Bzowski:2013sza,Bzowski:2015pba} \footnote{In position space one could get rid of divergences by  working at non-coincident points, but in momentum space one cannot do this and this leads to UV divergences.}. The story of spinning correlators is even more complicated \cite{Bzowski:2013sza,Bonora:2015nqa,Bzowski:2017poo,Bzowski:2018fql}.  




Parity-odd correlation functions in momentum space have received very little attention. In three-dimensions they appear naturally. Consider, for example, the free fermion theory in three-dimensions. The scalar primary operator with the lowest dimension is given by ${\bar \psi}\psi$, and it is odd under parity. A correlator with an odd number of insertions of this operator will be parity-odd. Another place where parity-odd structures arise naturally is CFTs with a broken parity. A prime example of such CFTs is Chern-Simons matter theories \cite{Aharony:2011jz,Giombi:2011kc,Maldacena:2012sf,Aharony:2012nh,GurAri:2012is,Giombi:2016zwa,Aharony:2018pjn,Kalloor:2019xjb}. In \cite{Maldacena:2012sf}, it was argued that in such theories :
\begin{equation}
\langle J_{s_1} J_{s_2} J_{s_3}\rangle =\alpha \langle J_{s_1} J_{s_2} J_{s_3}\rangle_{\text{Free-Boson}}+\beta \langle J_{s_1} J_{s_2} J_{s_3}\rangle_{\text{Free-Fermion}}+\gamma \langle J_{s_1} J_{s_2} J_{s_3}\rangle_{\text{odd}}
\end{equation} 
where $J_s$ is the spin $s$ conserved current, the subscript odd indicates a parity-odd contribution to the correlator and $\alpha,\beta,\gamma$ are theory dependent constants. In three-dimensions \footnote{In four dimensions, the 3-point function of stress-tensor is given by \cite{Osborn:1993cr}
$$\langle TTT\rangle=a \langle TTT\rangle_{\text{FB}}+b\langle TTT\rangle_{\text{FF}}+c\langle TTT\rangle_{\text{Maxwell}}.$$} we cannot obtain the parity-odd part of the correlator from the free theory and we need to use CFT techniques. 

There are other instances where parity-odd contribution might be interesting. For example, although parity-odd correlators are as yet unobserved, the CMB bispectrum - a measure of non-gaussianity - could contain parity-odd contributions. Such a contribution can arise, for example, from an inflationary action with higher derivative corrections such as cubic Weyl tensor terms \footnote{These are of the form $\int \widetilde{W} W^2$ where $W$ denotes the Weyl tensor and $\widetilde{W}$ its Hodge-dual.}.  Such terms could source contributions to the primordial graviton bispectrum, as first discussed in \cite{Maldacena:2011nz} using the spinor-helicity technique. In \cite{Soda:2011am} it was shown that the parity-violating (odd) contribution in the non-gaussianity of the primordial gravitational waves CMB vanishes in an exact de-Sitter background, but exists in inflationary quasi de-Sitter where it is proportional to the slow-roll parameter.  
Other studies of parity-violating CMB bispectrum include \cite{Lue:1998mq, Shiraishi:2014roa, Shiraishi:2014ila}. Our results on the parity-odd structures for various 3-point functions constrain the momentum space form of any parity-violating bispectrum that exists.

In this paper, we use two complementary approaches to determine the momentum space structure of parity-odd CFT 3-point functions. The first method is the more direct one and involves solving momentum-space Ward identities. This approach was developed and used for the parity-even sector in \cite{Bzowski:2013sza}.

The second approach utilises spin-raising and weight-shifting operators. These have been used in the conformal bootstrap literature \cite{Costa:2011dw,Costa:2011mg,Karateev:2017jgd} and more recently in fixing the form of cosmological correlators \cite{Baumann:2019oyu,Baumann:2020dch}. We will construct parity-odd spin-raising and weight-shifting operators in momentum space and use them on scalar seed correlators to generate parity-odd spinning correlators.


The rest of this paper is organised as follows. In Section \ref{3app}, besides setting up the notation and terminology, we outline the two different techniques that we use in this paper to determine parity-odd 3-point functions. We also briefly discuss the divergences that arise, and their regularisation. In Section \ref{dim}, we give an overview of the possible parity-odd 3-point structures in embedding space in various dimensions. In Section \ref{3ptCWI}, we use momentum space conformal Ward identities to fix the parity-odd part of $\langle JJO\rangle$. In Section \ref{weight}, we construct parity-odd spin-raising and weight-shifting operators in momentum space and use them to determine spinning correlators by their action on simple scalar seed correlators. We show that the results for $\langle JJO\rangle$ obtained using spin-raising and weight-shifting operators match the results for the same obtained in Section \ref{3ptCWI}. In Section \ref{JJJsection}, we compute the parity-odd 3-point function of the spin-one conserved current using weight-shifting and spin-raising operators. In Section \ref{TTO}, we compute the form of the odd part of $\langle TTO_{\Delta} \rangle$ for chosen conformal dimensions of $O$. In the free fermion theory, where $\Delta=2$, we match the results obtained using spin-raising and weight-shifting operators with the answer obtained by an explicit computation in the free theory. In Section \ref{4dsection}, we construct parity-odd spin-raising and weight-shifting operators in four-dimensions and use them to construct the parity-odd part of the non-trivial correlator $\langle JJJ\rangle$ in four dimensions.
We conclude with a discussion in Section \ref{FDir}. In the appendices we elaborate on various technical details. In Appendix \ref{embeddingspaceapp}, we review essential details on the embedding space formalism. In Appendix \ref{2pt},  we discuss the basics of momentum space two-point functions. In Appendix \ref{schouten}, we present the Schouten identities relevant to us. 
In Appendix \ref{CD}, we give some computational details of the results presented in Section \ref{TTO}. Appendix \ref{ParityEvenOperators} gives the form of various parity-even spin-raising and weight-shifting operators. In Appendix \ref{TTT}, we argue on grounds of permutation symmetry that certain 3-point correlators with spinning operators in four-dimensions vanish.


\section{Two approaches to determining momentum space correlators}\label{3app}
	Determining correlation functions is a significantly harder task in momentum space than in position space. For parity-odd correlators this gets even more tedious. We will now discuss two different approaches to determining momentum space correlators. We also discuss certain subtleties and limitations associated with the two approaches.
	
	\subsection{Using Conformal Ward identity : Strategy} 
	In the first approach, following \cite{Coriano:2013jba,Bzowski:2013sza} and \cite{Bzowski:2015pba,Bzowski:2017poo,Coriano:2018bbe,Bzowski:2018fql,Coriano:2018tgn} where parity-even 3-point functions were determined, we start with an ansatz of the form $\sum_m A_m(k_i) \mathcal{T}_m$ for the correlator. Here $\mathcal{T}_m$ are all possible tensor structures that are allowed by symmetry and $A_m$ are form factors which are functions of the momenta magnitudes ($k_i$). The form factors are constrained by permutation symmetries (if any) of the correlator and by momentum space Ward identities. The latter lead to partial differential equations which can then be solved to determine the form factors, up to undetermined constants that depend on the specific theory. In Section \ref{3ptCWI}, we use this method to fix the {\it parity-odd} part of certain 3-point functions in momentum space. An excellent mathematica package that we found useful in these computations is \cite{Bzowski:2020lip}.

Let us now describe the momentum space Ward identities associated with dilatation symmetry and special conformal transformations.

\subsection{Dilatation and Special Conformal Ward identities}\label{Skndris1}

	We will denote the $n$-point Euclidean correlation function of primary operators $\mathcal O_1,\ldots,\mathcal O_n$ by $\langle\mathcal O_1(\boldsymbol{k}_1)\ldots \mathcal O_n( \boldsymbol{k}_n)\rangle$. We suppress the Lorentz indices of the operators for brevity. The correlator with the momentum conserving delta function stripped off is denoted as :
\begin{align}
\langle\,\mathcal O_1( \boldsymbol{k}_1)\,\ldots\,\mathcal O_n( \boldsymbol{k}_n)\,\rangle\equiv(2\pi)^d\delta^{(3)}( \boldsymbol{k}_1+\ldots+ \boldsymbol{k}_n)\langle\langle\,\mathcal O_1( \boldsymbol{k}_1)\,\ldots\,\mathcal O_n( \boldsymbol{k}_n)\,\rangle\rangle\,.
\end{align}
An $n$-point correlator with scalar or spinning operator insertions satisfies the following dilatation Ward identity \cite{Bzowski:2013sza} :
\begin{align}
0=\left[-(n-1)d+\sum_{j=1}^n\,\Delta_j-\sum_{j=1}^{n-1}\,k_j^\alpha\,\frac{\partial}{\partial k_j^\alpha}\right]\langle\langle\,\mathcal O_1( \boldsymbol{k}_1)\,\ldots\,\mathcal O_n( \boldsymbol{k}_n)\,\rangle\rangle\,.
\end{align}
This constrains the correlator to have the following scaling behaviour :
\begin{align}
\langle\langle\,\mathcal O_1(\lambda\,\boldsymbol{k}_1)\,\ldots\,\mathcal O_n(\lambda\,\boldsymbol{k}_n)\,\rangle\rangle&=\lambda^{-\left[(n-1)d-\sum_{i=1}^n\Delta_i\right]}\langle\langle\,\mathcal O_1(\boldsymbol{k}_1)\,\ldots\,\mathcal O_n(\boldsymbol{k}_n)\,\rangle\rangle\,.
\end{align} 
The special conformal Ward identity on an $n$-point correlator with both scalar and spinning operators is \cite{Bzowski:2013sza} :
\begin{align}
\label{Kkappexplicit}
0&=\sum_{j=1}^{n-1}\left[2(\Delta_j-d)\frac{\partial}{\partial k_{j}^{\kappa}}-2k_j^\alpha
\frac{\partial}{\partial k_{j}^{\alpha}}\frac{\partial}{\partial k_{j}^{\kappa}}+k_j^\kappa\frac{\partial}{\partial k_{j}^{\alpha}}\frac{\partial}{\partial k_{j\alpha}}\right]\langle\langle\,\mathcal O_1(\boldsymbol{k}_1)\,\ldots\,\mathcal O_n(\boldsymbol{k}_n)\,\rangle\rangle\cr
&\hspace{.2cm}+2\sum_{j=1}^{n-1}\,\sum_{k=1}^{n_j}\,\left(\delta^{\mu_{jk}\kappa}\,\frac{\partial}{\partial k_j^{\alpha_{jk}}}-\delta^\kappa_{\alpha_{jk}}\frac{\partial}{\partial k_{j_{\mu_{jk}}}}\right)\langle\langle\,\mathcal O_1^{\mu_{11}\ldots\mu_{1r_1}}( \boldsymbol{k}_1)\,\ldots \mathcal O_j^{\mu_{j1}\ldots\alpha_{jk}\ldots\mu_{jr_j}}( \boldsymbol{k}_j)\,\ldots \mathcal O_n^{\mu_{n1}\ldots\mu_{nr_n}}( \boldsymbol{k}_n)\,\rangle\rangle\cr
\end{align}
In the second line of the RHS of the above equation, the indices of the generator mix with the spin indices of the correlator. In principle, one can solve this equation and get the desired correlator \cite{Bzowski:2013sza}. However, for parity-odd structures in three-dimensions, the computation gets complicated and has not yet been done.  

We will always be working with correlation functions with the momentum conserving delta function stripped off. From here on we will drop the double angular brackets notation to avoid clutter and use single angular brackets everywhere.
	\subsubsection{Divergences}
Triple-$K$ integrals arise as solutions to primary conformal Ward identities which are second order differential equations \cite{Bzowski:2013sza}. Along with the three momenta, they are expressed in terms of four other parameters :
\begin{align}
\label{generaltriplek}
I_{\alpha\{\beta_1\beta_2\beta_3\}}(k_1,k_2,k_3)\equiv\int_0^{\infty}dx\; x^{\alpha}\prod_{j=1}^3k_j^{\beta_j}K_{\beta_j}(k_jx)
\end{align}
where $K_{\beta_j}$ is a modified Bessel function of the second kind. 
While the integral is well behaved at its upper limit, it is convergent at $x=0$ only if \cite{Bzowski:2013sza,Bzowski:2017poo} :
\begin{align}
\label{triplkconditions}
\alpha+1-|\beta_1|-|\beta_2|-|\beta_3|>0
\end{align}
When the integral is divergent one can regulate it using two parameters $u$ and $v$ \cite{Bzowski:2013sza,Bzowski:2017poo} :
\begin{align}
I_{\alpha\{\beta_1\beta_2\beta_3\}}\rightarrow I_{\alpha+u\epsilon\{\beta_1+v\epsilon,\beta_2+v\epsilon,\beta_3+v\epsilon\}}
\end{align}
The regularised triple-$K$ integral is convergent except when \cite{Bzowski:2013sza,Bzowski:2017poo} :
%
\begin{align}
\label{regtriplkconditions}
\alpha+1\pm\beta_1\pm\beta_2\pm\beta_3=-2n,\quad\quad n\in\mathbb Z_{\ge 0}
\end{align}
for any choice of signs.  When \eqref{regtriplkconditions} is satisfied, the integral is singular in the regulator $\epsilon$ and we will denote the divergence by the choice of signs $(\pm\pm\pm)$ for which \eqref{regtriplkconditions} is satisfied. 

Divergences of the type $(---)$ are called ultra-local and they occur when all the three operators are co-incident in position space. In momentum space, this manifests as the divergent term being analytic in all three momenta squared. Such divergences must, in general, be removed using counter-terms that are cubic in the sources, and they give rise to conformal anomalies.

Divergences of the type $(--+)$ and its permutations are called semi-local divergences. In position space, this is a divergence that occurs when two of the operators in the correlator are at co-incident points. In momentum space, the divergence is said to be semi-local when the $O(1/\epsilon)$ term is analytic in any two of the three momenta squared. In general, these divergences must be removed by counter-terms that have two sources and an operator. Such terms lead to non-trivial beta functions.

Divergences of the kind $(+++)$ and $(++-)$ are non-local and they occur even when all three operators are at separated points in position space. In momentum space, such a divergence is analytic in at most one of the momenta squared. This is not a physical divergence and arises because the triple-$K$ integral representation of the correlator is singular. In this case no counter-term exists and the divergence is removed by imposing the condition that the constant multiplying the triple-$K$ integral vanishes as an appropriate power of $\epsilon$.

\subsubsection{Counter-terms}
As we discussed above, divergences of the kind $(--+)$ and $(---)$ that correspond to ultra-local and semi-local divergences are removed using suitable counter-terms. In the case of parity-even correlators this has been extensively studied in  \cite{Bzowski:2013sza,Bzowski:2015pba,Bzowski:2017poo,Bzowski:2018fql}.

We will now list a few potential counter-terms that could turn out to be useful in our study of parity-odd correlators. For ultra-local divergences, for example we have : 
%
%
\begin{align}
\label{ctultralocal}
\int d^3x\,F_3(A)\,\Box^n\phi,\quad\int d^3x\,C_{\mu\nu}\,R^{\mu\nu}\,\Box^n\phi,\quad
\int d^3x\,C_{\mu\nu}\,R\,\nabla^\mu\,\nabla^\nu\,\Box^n\phi
\end{align}
and for semi-local divergences :
\begin{align}
\label{ctsemilocal}
\int d^{3}x\,\epsilon^{\mu\nu\lambda}\,F_{\mu\nu}\,J_{\lambda}\,\Box^n\phi,\quad 
\int d^3x\,A^\mu\,J_\mu\,\Box^n\phi,\quad \int d^3x\,F^{\mu\nu}J_\mu D_\nu\phi,\quad \int d^3x\,C_{\mu\nu}\,T^{\mu\nu}\,\Box^n\phi
\end{align}
where $F_3(A)$ is the Chern-Simons form in three-dimensions given by,
\begin{equation}
F_3(A)=\epsilon_{\mu\nu\lambda}\left(A^{\mu}_a\partial^{\nu}A^{\lambda}_a+\frac{2}{3}f^{abc}A^{\mu}_aA^{\nu}_bA^{\lambda}_c\right)\,,
\end{equation}
$C_{\mu\nu}$ is the Cotton-York tensor given by,
\begin{align}
C_{\mu\nu}=\nabla^{\rho}\left(R^{\sigma}_{\mu}-\frac 14 R\,g^{\sigma}_{\mu}\right)\epsilon_{\rho\sigma\nu}\,,
\end{align}
and $R_{\mu\nu}$ and $R$ are the Ricci tensor and the Ricci scalar respectively.

In the above list of possible counter-terms \eqref{ctsemilocal} we have included certain parity-even terms such as $\int d^3x\,A^\mu\,J_\mu\,\Box^n\phi$ and $\int d^3x\,F^{\mu\nu}J_\mu D_\nu\phi$. These counter-terms could give rise to the 2-point function of currents, which has a parity-odd contribution $\langle J^\mu(p)J^\nu(-p)\rangle\propto\epsilon^{\mu\nu\rho}p_\rho$.
	\subsection{Using Weight-shifting and Spin-raising operators}
	The second method of computing correlation functions in momentum space hinges on the technique of weight-shifting and spin-raising operators. In position space, this technique was initiated in \cite{Costa:2011dw} and extensively developed in \cite{Karateev:2017jgd}. In this approach, starting from certain seed correlators, the action of conformally covariant weight-shifting and spin-raising operators generates the desired correlator.
	\par
	To describe this method in some detail, let us consider a spinning correlator $\langle J_{s_1}J_{s_2}J_{s_3}\rangle.$  The first step is to count the number of independent tensor structures associated with this correlator.  For parity-even correlators this number in position space is given by \cite{Costa:2011mg} :
	\begin{equation}
	N_{3d}^+(l_1, l_2, l_3)=2 l_{1} l_{2}+l_{1}+l_{2}+1-\frac{p(p+1)}{2}
	\end{equation}
	where $p=\text{max}(0, l_1+l_2-l_3)$.
	The second step is to consider a seed correlator of the form $\langle O_{\Delta_1}O_{\Delta_2}J_{s_3}\rangle$ and find out $N_{3d}^+(l_1, l_2, l_3)$ ways to reach $\langle J_{s_1}J_{s_2}J_{s_3}\rangle.$ This involves acting upon the seed correlator with various spin-raising and weight-shifting operators. 
	
	In momentum space, we are constrained in our choice of seed correlators because correlators of the form $\langle J^{(l)}O_{\Delta_1}O_{\Delta_2} \rangle$, where $J^{(l)}$ is a spin-$l$ conserved current, are non-zero only when $\Delta_1=\Delta_2$. A more convenient approach was recently advocated in \cite{Baumann:2019oyu,Baumann:2020dch} to compute (parity-even) spinning cosmological correlators
	where instead of starting from the seed $\langle O_{\Delta_1}O_{\Delta_2}J_{s_3}\rangle,$ one starts from $\langle O_{\Delta_1}O_{\Delta_2}O_{\Delta_3} \rangle$, and apply spin-raising and weight-shifting operators such that the resulting correlator saturates the Ward-Takahashi identity. See Section 4.2.2 of \cite{Baumann:2020dch} for an example.

	\par
\subsubsection{Subtleties with the weight-shifting and spin-raising operator approach}
	In momentum space, one must consider the types of divergences in the seed and target correlators. It is not always possible to reach a target correlator starting from a seed correlator although a naive application of the spin-raising and weight-shifting operators might suggest so. This is most easily understood in the case of scalar correlators. As a concrete example of such a situation, consider the following two correlators in three-dimensions :
	\begin{align}
	\langle O_1(k_1)O_1(k_2)O_2(k_3) \rangle&=\frac{1}{k_1 k_2}\\[6 pt]
	\langle O_2(k_1)O_2(k_2)O_2(k_3) \rangle&=-\text{log}\left(\frac{k_1+k_2+k_3}{\mu}\right)
	\end{align}
	where $\mu$ is the renormalisation scale. Although it might seem like we can use the weight-shifting operator $W_{12}^{++}$ (defined in \eqref{W12plus}) to go from the first correlator to the second, this is clearly not possible as $\langle O_2(k_1)O_2(k_2)O_2(k_3) \rangle$ violates scale invariance whereas the seed correlator $\langle O_1(k_1)O_1(k_2)O_2(k_3) \rangle$ does not, i.e.
	\begin{equation}
	W_{12}^{++}\langle O_1(k_1)O_1(k_2)O_2(k_3) \rangle \neq \langle O_2(k_1)O_2(k_2)O_2(k_3) \rangle
	\end{equation}
	The above example tells us that weight-shifting operators fail to reproduce the correct correlators when the divergence type changes from non-local to semi-local or ultra-local. The conditions for various types of divergences, in terms of scaling dimensions of the operator insertions, are given by :
	\begin{center}
		\begin{tabular}{|c | c |}
			\hline
			$(---)$ & $\Delta_1+\Delta_2+\Delta_3=2d+2k_1$\\
			\hline
			$(--+)$ & $\Delta_1+\Delta_2-\Delta_3=d+2k_2$\\
			\hline
			$(++-)$ & $-\Delta_1-\Delta_2+\Delta_3=2k_3$\\
			\hline
			$(+++)$ & $\Delta_1+\Delta_2+\Delta_3=d-2k_4$\\
			\hline
		\end{tabular}
	\end{center}
	where $k_1, k_2, k_3, k_4 \geq 0$. We can see that the only time the divergence structure changes is when $k_i=0$. For the non-local cases in three-dimensions, these correspond to the following for the seed correlator :
	\begin{align}
	\begin{split}
	\Delta_3&=\Delta_1+\Delta_2
	\end{split}
	\begin{split}
	(++-)
	\end{split}\\[6 pt]
	\begin{split}
	\Delta_3&=3-\Delta_1-\Delta_2
	\end{split}
	\begin{split}
	(+++)
	\end{split}
	\end{align}
	When either of these conditions is satisfied by the seed correlator, the action of $W_{12}^{++}$ does not reproduce the correct result. However, $W_{12}^{--}$ works as it can be checked that it does not change the type of divergence. 
	\par

	\section{Allowed parity-odd 3-point functions in various dimensions}\label{dim}

Parity-odd structures for three-point correlators can exist in simple theories such as the free fermion theory in three-dimensions and in CFTs which do not have a parity symmetry. Such structures change sign under inversion and they always involve the antisymmetric epsilon tensor. The existence, or non-existence, of such correlators in various dimensions is easily seen in the embedding space formalism \cite{Costa:2011mg, Costa:2011dw}. In embedding space, $d$-dimensional odd-correlators are characterised by $(d+2)$-dimensional epsilon tensor. Although we restrict our attention to correlators involving spin-zero, spin-one and spin-two currents, the analysis below can be easily generalised to parity-odd correlators involving traceless symmetric operators in various dimensions.  We refer the reader to Appendix \ref{embeddingspaceapp} for some details on the embedding space formalism.
   
   
One of the constraints on correlators with a spinning operator is that they are transverse, i.e.  the epsilon structure should be invariant under $Z_i \to Z_i + \beta X_i$. This constrains the possible epsilon structures one can have in a given dimension.
\subsection{Three-dimensions}
In three-dimensions, transversality implies that the parity-odd invariants that can exist are:
\begin{align}
\epsilon(Z_1Z_2X_1X_2X_3), \quad \epsilon(Z_2Z_3X_1X_2X_3), \quad \epsilon(Z_3Z_1X_1X_2X_3)
\end{align}
 This immediately implies that the following parity-odd correlation function is zero :
\begin{equation}
\langle J_s O_{\Delta_1} O_{\Delta_2}\rangle_{\text{odd}}=0
\end{equation} where $O_{\Delta_1},O_{\Delta_2}$ are scalar operators with dimensions as indicated and $J_s$ is a spin $s$ operator.
However, correlators of the form  $\langle J_{s_1} J_{s_2}O_\Delta\rangle_\text{{odd}}$ and $\langle J_{s_1} J_{s_2}J_{s_3}\rangle_\text{{odd}}$ are non-zero. See \cite{Giombi:2011rz} for details.
In Section \ref{jooodd0}, we will explicitly show that, in momentum space :
\begin{equation}
\langle J_{\mu} O O\rangle_{\text{odd}}=0.
\end{equation}
We will also calculate other parity-odd three-point functions in subsequent sections.

\subsection{Four-dimensions}\label{4dpodd}
In four-dimensions, transversality allows only the following parity-odd invariant :
\begin{align}
\epsilon(Z_1Z_2Z_3X_1X_2X_3)
\end{align}
This implies that in four-dimensions the following correlators are zero :
\begin{equation}
\langle J_{s_1} O O\rangle_{\text{odd}} =0,~~\langle J_{s_1} J_{s_2}O\rangle_{\text{odd}}=0
\end{equation}
Correlators are further constrained by symmetry requirements. For example,  
\begin{equation}
\langle JJT\rangle_\text{{odd}}=0, ~~ \langle TTT\rangle_\text{{odd}}=0. 
\end{equation}
See Appendix \ref{TTT} for details. While these correlators are zero, $\langle J^a_{\mu}J^b_{\nu}J^c_{\rho}\rangle$ can be non-zero.
\subsection{Five-dimensions and above}
In five-dimensions, the only parity-odd invariant allowed by transversality is :
\begin{align}
\epsilon(Z_1Z_2Z_3X_1X_2X_3X_4)
\end{align}
The structure of the contracted epsilon tensor makes it clear that we cannot have any odd three-point function in five-dimensions 
\footnote{However, the above odd structure hints at the possibility of odd four-point correlators in five-dimensions. Since there exists no parity-odd three-point function in five-dimensions, the parity-odd four point function will be a contact term.}. At this point it is also clear that we cannot have an odd three-point function in $d \ge 5$, or an odd four-point function in $d \geq 6$. 


\section{Momentum space parity-odd 3-point functions in three-dimensions using conformal Ward identity}\label{3ptCWI}
In this section, we will use the direct approach of solving the PDEs for the form factors which result from the momentum space conformal Ward identities. We will illustrate this using the correlators $\langle JOO \rangle $ and $\langle JJO \rangle $ as examples. This is an extension of the analysis in \cite{Bzowski:2013sza} to the parity-odd sector. We also discuss the divergences that can arise, and the counter-terms that regulate them. We conclude this section with a small discussion on the difficulties in using this method to compute correlators involving more general spinning operators.

\subsection{$\langle J^{\mu} O O \rangle_{\text{odd}}$}
\label{jooodd0}
Let us consider the parity-odd part of the $\langle J^{\mu} O O \rangle$ correlator where $J^{\mu}$ is a conserved current and $O$ has scaling dimension $2$. A suitable ansatz for the correlator is : 
\begin{equation}
\langle O^a(k_1) O^b(k_2) J^{\mu c}(k_3) \rangle_{\text{odd}}=A(k_1,k_2,k_3)\,f^{abc}\,\epsilon^{\mu k_1 k_2}
\end{equation}
Throughout this paper, we use notations such as $\epsilon^{\mu\nu k_1}$ and $\epsilon^{\mu k_1 k_2}$  and they stand for the epsilon tensor contracted with the momenta :
\begin{align}
\epsilon^{\mu\nu k_1}=\epsilon^{\mu\nu\rho}\,k_{1\rho},\quad\epsilon^{\mu k_1 k_2}=\epsilon^{\mu\nu\rho}\,k_{1\nu}\,k_{2\rho}
\end{align}
We have considered $J^{\mu}$ to be in the third position as this makes the action of $K^{\kappa}$ simpler \eqref{Kkappexplicit}. Without non-abelian indices the correlator is zero as $(k_1 \leftrightarrow k_2)$ exchange symmetry would require $A(k_1, k_2)=-A(k_2, k_1)$. After acting with $K^{\kappa}$, the primary Ward identities are given by :
\begin{align}\label{jooWI}
\begin{split}
\frac{\partial^2 A}{\partial k_1^2}-\frac{\partial^2 A}{\partial k_3^2}-\frac{2}{k_3}\frac{\partial A}{\partial k_3}&=0\\[6 pt]
\frac{\partial^2 A}{\partial k_2^2}-\frac{\partial^2 A}{\partial k_3^2}-\frac{2}{k_3}\frac{\partial A}{\partial k_3}&=0
\end{split}
\end{align}
The above differential equations can be solved in terms of triple-$K$ integrals \eqref{generaltriplek} to get :
\begin{equation}\label{jooform}
A(k_1,k_2,k_3)=c_1\,I_{\frac{3}{2}\{\frac{1}{2}\frac{1}{2}-\frac{1}{2}\}}=c_1\,\frac{1}{k_3(k_1+k_2+k_3)}
\end{equation}
The correlator also satisfies an independent secondary Ward identity given by :
\begin{equation}\label{joosec}
2\,A+\frac{k_1^2-k_2^2+k_3^2}{k_3}\frac{\partial A}{\partial k_3}+2\,k_1\frac{\partial A}{\partial k_1}=0
\end{equation}
The right hand side of the above equation is proportional to the 2-point function $\langle O_2(k) O_2(-k) \rangle$. However, the scalar two-point function has no parity-odd contribution and thus the R.H.S. of \eqref{joosec} is zero. Substituting \eqref{jooform} into \eqref{joosec} gives
\begin{equation}
c_1=0\implies A(k_1,k_2,k_3)=0
\end{equation}
Thus we conclude that :
\begin{equation}
\langle O^a(k_1) O^b(k_2) J^{\mu c}(k_3) \rangle_{\text{odd}}=0
\end{equation}
This result can be generalised to scalar operators of arbitrary scaling dimensions.

\subsection{$\langle J^{\mu} J^{\nu} O\rangle_{\text{odd}}$}
Here we will consider the parity-odd part of the correlator $\langle J^\mu J^\nu O\rangle$. We start with the following ansatz for the correlator  :
\begin{align}
\label{JJOansatz2}
\langle J^\mu(k_1)\,J^\nu(k_2)\,O(k_3)\rangle_{\text{odd}}=\pi^{\mu}_{\alpha}(k_1)\pi^{\nu}_{\beta}(k_2)\left[\widetilde A(k_1,k_2,k_3) \epsilon^{\alpha k_1 k_2} k_1^{\beta}+\widetilde B(k_1,k_2,k_3) \epsilon^{\beta k_1 k_2}k_2^{\alpha}\right]
\end{align}
where the orthogonal projector $\pi^\mu_\nu(p)$ is given by :
 \begin{align}
 \pi_{\mu}^{\nu}({p})&\equiv\delta_{\mu}^{\nu}-\frac{p^{\nu}\,p_{\mu}}{p^{2}}\,.\label{projk1}
 \end{align}
%
The ansatz \eqref{JJOansatz2} is chosen such that the correlator is transverse with respect to $k_1^{\mu}$ and $k_2^{\nu}$. Demanding symmetry under the exchange : $\left(k_1, \mu\right) \leftrightarrow \left(k_2, \nu\right)$ gives the following relation between the form factors :
\begin{equation}
\label{exchangesymm}
\widetilde A(k_1, k_2, k_3)=-\widetilde B(k_2, k_1, k_3)
\end{equation}
Using the definition of projectors \eqref{projk1}, the ansatz \eqref{JJOansatz2} expands to the following :
\begin{align}
\label{JJOansatz2expanded}
\langle J^\mu(k_1)\,J^\nu(k_2)\,O(k_3)\rangle_{\text{odd}}&=\widetilde A(k_1,k_2,k_3)\epsilon^{\mu k_1 k_3}\left[\frac{(k_1^{\nu}+k_3^{\nu})(k_1^2+k_1\cdot k_3)}{k_2^2}-k_1^{\nu}\right]\cr
&\hspace{.5cm}+\widetilde B(k_1,k_2,k_3)\epsilon^{\nu k_1 k_3}\left[(k_1^{\mu}+k_3^{\mu})-\frac{k_1^{\mu}(k_1^{2}+k_1\cdot k_3)}{k_1^2}\right]
\end{align}
where we have used momentum conservation to choose $k_1$ and $k_3$ as the independent momenta. 

We now use Schouten identities \eqref{jjoschouten1} and \eqref{jjoschouten2} to get rid of the $\epsilon^{\mu k_1 k_2}$ tensor structure and re-express the ansatz in \eqref{JJOansatz2expanded} as :
\begin{align}
\label{JJOansatz1}
\langle J^\mu(k_1)\,J^\nu(k_2)\,O(k_3)\rangle_{\text{odd}}=-\epsilon^{\nu k_1k_3}\left(Ak_1^{\mu}-Bk_1^{\mu}-Bk_3^{\mu}\right)-\left(\epsilon^{\mu\nu k_1}+\epsilon^{\mu\nu k_3}\right)\left(Ak_1^2+B(k_1\cdot k_2)\right)
\end{align}
where the new form factors $A(k_1,k_2,k_3)$ and $B(k_1,k_2,k_3)$ are given in terms of $\widetilde A(k_1,k_2,k_3)$ and $\widetilde B(k_1,k_2,k_3)$ as follows :
\begin{align}
A(k_1,k_2,k_3)&=\widetilde A(k_1,k_2,k_3)+\widetilde B(k_1,k_2,k_3)+\widetilde B(k_1,k_2,k_3)\,\frac{k_1\cdot k_3}{k_1^2}\cr
B(k_1,k_2,k_3)&=\widetilde B(k_1,k_2,k_3)-\widetilde A(k_1,k_2,k_3)\frac{k_1\cdot k_2}{k_2^2}
\end{align}
Note that the exchange symmetry \eqref{exchangesymm} continues to hold between $A$ and $B$: 
\begin{align}
A(k_1, k_2, k_3)=-B(k_2, k_1, k_3)
\end{align}
We will now obtain the primary and secondary Ward identities that $A(k_1,k_2,k_3)$ and $B(k_1,k_2,k_3)$ satisfy, by letting the generator of special conformal transformations $K^\kappa$ \eqref{Kkappexplicit} act on the ansatz \eqref{JJOansatz1} :
\begin{align}
K^\kappa\langle J^\mu(k_1) J^\nu(k_3) O(k_2)\rangle_{\text{odd}}&=\Big[-2\frac{\partial}{\partial k_1^\kappa}-2k_1^\alpha\frac{\partial}{\partial k_1^\alpha}\frac{\partial}{\partial k_1^\kappa}+k_{1,\kappa}\frac{\partial}{\partial k_1^\alpha}\frac{\partial}{\partial k_{1\alpha}}\cr
&\hspace{.7cm}+2(\Delta_3-3)\frac{\partial}{\partial k_3^\kappa}-2k_3^\alpha\frac{\partial}{\partial k_3^\alpha}\frac{\partial}{\partial k_3^\kappa}+k_{3,\kappa}\frac{\partial}{\partial k_3^\alpha}\frac{\partial}{\partial k_{3\alpha}}\Big]\langle J^\mu(k_1) J^\nu(k_2) O(k_3)\rangle\cr
&\hspace{.7cm}+2\left(\delta^{\mu\kappa}\frac{\partial}{\partial k_1^\alpha}-\delta^\kappa_\alpha\frac{\partial}{\partial k_{1,\mu}}\right)\langle J^\alpha(k_1) J^\nu(k_2) O(k_3)\rangle
\end{align}
Note that by choosing $k_1$ and $k_3$ as the independent momenta, we got rid of one set of terms in the generator $K^\kappa$ that mixes with the index structure of the correlator.

The primary Ward identities satisfied by $A(k_1,k_2,k_3)$ are given by :
\begin{align}
\label{PWI}
\frac{\partial^2 A}{\partial k_1^2}+\frac{\partial^2 A}{\partial k_3^2}+\frac{2k_1}{k_3}\frac{\partial^2 A}{\partial k_1\,\partial k_3}+\frac{2k_2}{k_3}\frac{\partial^2 A}{\partial k_2\,\partial k_3}+\frac{2}{k_1}\frac{\partial A}{\partial k_1}+\frac{8}{k_3}\frac{\partial A}{\partial k_3}=0\nonumber\\[15pt]
\frac{\partial^2 A}{\partial k_3^2}+\frac{\partial^2 A}{\partial k_2^2}+\frac{2k_1}{k_3}\frac{\partial^2 A}{\partial k_1\,\partial k_3}+\frac{2k_2}{k_3}\frac{\partial^2 A}{\partial k_2\,\partial k_3}+\frac{8}{k_3}\frac{\partial A}{\partial k_3}=0
\end{align}
Similarly, the equations for $B(k_1,k_2,k_3)$ are given by :
\begin{align}
\frac{\partial^2 B}{\partial k_2^2}+\frac{\partial^2 B}{\partial k_3^2}+\frac{2k_1}{k_3}\frac{\partial^2 B}{\partial k_1\,\partial k_3}+\frac{2k_2}{k_3}\frac{\partial^2 B}{\partial k_2\,\partial k_3}+\frac{2}{k_2}\frac{\partial B}{\partial k_2}+\frac{8}{k_3}\frac{\partial B}{\partial k_3}=0\nonumber\\[15pt]
\frac{\partial^2 B}{\partial k_3^2}+\frac{\partial^2 B}{\partial k_1^2}+\frac{2k_1}{k_3}\frac{\partial^2 B}{\partial k_1\,\partial k_3}+\frac{2k_2}{k_3}\frac{\partial^2 B}{\partial k_2\,\partial k_3}+\frac{8}{k_3}\frac{\partial B}{\partial k_3}=0
\end{align}
The general solution to both the primary Ward identities can be found in terms of triple-$K$ integrals \eqref{generaltriplek}. 
We solve for $\beta_1, \beta_2, \beta_3$ by substituting the triple-$K$ integral into the primary Ward identities, and obtain :
\begin{align}
A&\propto I_{\alpha\,\{\Delta_1-\frac d2 -1,\Delta_2-\frac d2,\Delta_3-\frac d2\}}\cr
B&\propto  I_{\alpha\,\{\Delta_1-\frac d2,\Delta_2-\frac d2-1,\Delta_3-\frac d2\}}
\end{align}
%
The unknown $\alpha$ is determined using the dilatation Ward identity. The action of the dilatation Ward identity on the ansatz gives the degree of the form factors :
\begin{align}
\label{degABJJO}
\text{deg}(A)&=\Delta_1+\Delta_2+\Delta_3-2d-N_A \cr
\text{deg}(B)&=\Delta_1+\Delta_2+\Delta_3-2d-N_B
\end{align}
where $N_A$ and $N_B$ are the tensorial dimensions of $A$ and $B$, defined as the number of momenta that multiply the form factor in the ansatz. We see from  \eqref{JJOansatz1} and \eqref{JJOansatz2} that $N_A=N_B=3$. Similarly, we impose the dilatation Ward identity on the triple-$K$ integral and get : 
\begin{equation}
\text{deg}(I_{\alpha\{\beta_j\}})=\beta_1+\beta_2+\beta_3-\alpha-1
\end{equation}
This must equal the degree of the form factors $A$ and $B$ \eqref{degABJJO} giving us :
\begin{equation}
\label{alphajjo}
\alpha=\sum_{i=1}^3\,(\beta_i-\Delta_i)+2d+2
\end{equation}
Thus we obtain : 
\begin{align}
\label{ABJJO}
\begin{split}
A&=c_1\,I_{\sum_{i=1}^3\,(\beta_i-\Delta_i)+2d+2\{\Delta_1-\frac d2 -1,\Delta_2-\frac d2,\Delta_3-\frac d2\}} \\
B&=c_2\,I_{\sum_{i=1}^3\,(\beta_i-\Delta_i)+2d+2\{\Delta_1-\frac d2,\Delta_2-\frac d2-1,\Delta_3-\frac d2\}}
\end{split}
\end{align}
where $c_1$ and $c_2$ are undetermined constants. 

We now present the explicit expressions for the two form factors for a few values of the scaling dimension of the scalar operator $O$. When the scalar operator has $\Delta_3=1$, we have,
\begin{align}
\label{JJOdelta1}
\begin{split}
A(k_1,k_2,k_3)=c_1\sqrt{\frac{\pi^3}{8}}\frac{1}{k_1k_3\left(k_1+k_2+k_3\right)^2},\cr
B(k_1,k_2,k_3)=c_2\sqrt{\frac{\pi^3}{8}}\frac{1}{k_2k_3\left(k_1+k_2+k_3\right)^2}
\end{split}
\end{align}
For $\Delta_3=2$ :
\begin{align}
\label{JJOdelta2}
\begin{split}
A(k_1,k_2,k_3)&=c_1\sqrt{\frac{\pi^3}{8}}\frac{1}{k_1\left(k_1+k_2+k_3\right)^2},\cr
B(k_1,k_2,k_3)&=c_2\sqrt{\frac{\pi^3}{8}}\frac{1}{k_2\left(k_1+k_2+k_3\right)^2}.
\end{split}
\end{align}
When $\Delta_3=3$ :
\begin{align}
\label{JJOdelta3}
\begin{split}
A(k_1,k_2,k_3)&=c_1\sqrt{\frac{\pi^3}{8}}\frac{k_1+k_2+2k_3}{k_1\left(k_1+k_2+k_3\right)^2} \\
B(k_1,k_2,k_3)&=c_2\sqrt{\frac{\pi^3}{8}}\frac{k_1+k_2+2k_3}{k_2\left(k_1+k_2+k_3\right)^2}.
\end{split}
\end{align}
We will now look at the secondary Ward identities to fix the undetermined constants $c_1$ and $c_2$ in \eqref{ABJJO}.

There is one independent secondary Ward identity in this case which leaves just one independent, undetermined constant. The identity is given by :
\begin{align}
\frac{k_1^2}{k_2}\frac{\partial A}{\partial k_2}+k_1\frac{\partial B}{\partial k_1}=0
\end{align}
Substituting the solutions for the form factors from \eqref{ABJJO} in this equation we get:
\begin{equation}\label{JJOsec}
c_2=-c_1
\end{equation}
which is exactly what is expected from symmetry considerations.

\subsubsection{Divergences and Renormalization}\label{Divergences}
We saw in equations \eqref{JJOdelta1}, \eqref{JJOdelta2} and \eqref{JJOdelta3} that the triple-$K$ integral is convergent for $\Delta=1, 2, 3$. For $\Delta_3>3$, the integral is singular in the regulator and in some cases, we will require counter-terms to remove this divergence. 

The generating functional for the theory is defined as :
\begin{equation}
Z=\int{D\phi\; \text{exp}\left(-\int{d^3x\; (S_{\phi}[A_{\mu}, g^{\mu\nu}]+\sqrt{g}O\phi+J^{\mu}A_{\mu}})\right)}
\end{equation}
where $\phi$ and $A_{\mu}$ are sources of the scalar operator and the conserved spin-one current respectively. For certain classes of divergences, the generating functional is modified by counter-terms. We classify the values of $\Delta_3$ into two classes based on the kinds of divergences that occur.

{$\bm{\Delta_3=4+2n}$} : When $\Delta_3=4+2n$ where $n \in \{0, 1 ,2,..\}$, \eqref{regtriplkconditions} is satisfied for the choice of signs given by $(+--)$ for $n=0$, i.e., $\Delta_3=4$. When $n > 0$, it is satisfied for the choice of signs $(+--)$ and $(-+-)$. 

To remove this singularity, we look at the following parity-odd counter-term from \eqref{ctsemilocal} 
\begin{equation}
\label{ctjjotype1}
S_{ct}=\int{d^{3+\epsilon}x\; \mu^{-\epsilon}\,\epsilon^{\mu\nu\lambda}\,F_{\mu\nu}\,J_{\lambda}\,\Box^{n}\phi}
\end{equation}
where $\mu$ is the renormalization scale. After taking suitable functional derivatives, the contribution to the correlator from this counter-term is given by 
\begin{align}
\langle J^{\mu}(x_1)J^{\nu}(x_2)O(x_3)\rangle_{ct}&=-a(\epsilon)\bigg[\Box^n\left(\delta^3(x_2-x_3)\epsilon^{\rho\nu\lambda}\partial_{2\rho}\langle J^{\lambda}(x_1)J^{\mu}(x_3) \rangle\right)\bigg.\cr
&\hspace{1.7cm}\bigg.-\Box^n\left(\delta^3(x_1-x_3)\epsilon^{\rho\mu\lambda}\partial_{1\rho}\langle J^{\lambda}(x_3)J^{\nu}(x_2) \rangle\right)\bigg]
\end{align}
A Fourier transform of the above gives :
\begin{align}
\label{jjocounterterm}
\langle J^{\mu}(k_1)J^{\nu}(k_2)O(k_3)\rangle_{ct}&=-a(\epsilon)\bigg(k_2^{2n} \epsilon^{\nu k_2\lambda}\pi^{\mu}_{\lambda}(k_1)\,k_1-k_1^{2n}\epsilon^{\mu k_1\lambda}\pi^{\nu}_{\lambda}(k_2)\,k_2\bigg)\nonumber\\[5 pt]
&=-a(\epsilon)\bigg[k_2^{2n}k_1\left(\epsilon^{\mu\nu k_2}+\frac{\epsilon^{\nu k_1 k_2}k_1^{\mu}}{k_1^2}\right)-k_1^{2n}k_2\left(\epsilon^{\mu\nu k_1}+\frac{\epsilon^{\mu k_1 k_2}k_2^{\nu}}{k_2^2}\right)\bigg]\mu^{-\epsilon}
\end{align}
where we used the following 2-point function \footnote{The counter-term that we used \eqref{ctjjotype1} could also contribute to the parity-even part of $\langle JJO\rangle$ since  the $\langle JJ\rangle $ 2-point-function has a parity-odd contribution.} :
\begin{align}
\langle J^{\mu}(k)J_{\nu}(-k) \rangle=\pi^{\mu}_{\nu}(k)\,k 
\end{align}
Using Schouten identities \eqref{jjoschouten1} and \eqref{jjoschouten2}, the ansatz for the correlator can be written as 
\begin{equation}
\label{jjonewansatz}
\langle J^{\mu}(k_1)J^{\nu}(k_2)O(k_3) \rangle=A_1\left(\epsilon^{\mu\nu k_2}k_1^2+\epsilon^{\nu k_1 k_2}k_1^{\mu}\right)+A_2\left(\epsilon^{\mu\nu k_1}k_2^2+\epsilon^{\mu k_1 k_2}k_2^{\nu}\right)
\end{equation}
When $\Delta_3=4$ the singular part of the regularised form factors are given by
\begin{align}
\label{A1A2jjosingular}
\begin{split}
A_1(k_1, k_2, k_3)=\frac{1}{k_1\epsilon},\quad A_2(k_1, k_2, k_3)=-\frac{1}{k_2\epsilon}
\end{split}
\end{align}
The contribution of the counter-term \eqref{jjocounterterm} to the correlator in this case ($\Delta_3=4$, or equivalently $n=0$) is given by :
\begin{align}
\label{jjo4counterterm}
\langle J^{\mu}(k_1)J^{\nu}(k_2)O(k_3)\rangle_{ct}=-a(\epsilon)\bigg[k_1\left(\epsilon^{\mu\nu k_2}+\frac{\epsilon^{\nu k_1 k_2}k_1^{\mu}}{k_1^2}\right)-k_2\left(\epsilon^{\mu\nu k_1}+\frac{\epsilon^{\mu k_1 k_2}k_2^{\nu}}{k_2^2}\right)\bigg]\mu^{-\epsilon}
\end{align}
Comparing \eqref{jjo4counterterm} and \eqref{jjonewansatz} along with \eqref{A1A2jjosingular} we see that choosing $a(\epsilon)=1/\epsilon$ cancels the singular part of the correlator. After removing the divergences, the resulting form factor is given by :
\begin{equation}\label{jjo4form}
A_1(k_1, k_2, k_3)=c_1\frac{3}{k_1}\text{log}\left(\frac{k_1+k_2+k_3}{\mu}\right)-c_1\frac{k_3^2+3k_3(k_1+k_2+k_3)}{k_1(k_1+k_2+k_3)^2}
\end{equation}
The second form factor is obtained by the following exchange :
\begin{align} 
A_2(k_1,k_2,k_3)=-A_1(k_2,k_1,k_3)
\end{align}
The anomalous dilatation Ward identity takes the form :
\begin{align}
\mu\frac{\partial A_1}{\partial \mu}=-\frac{c_1}{k_1}
\end{align}
%

$\bm{\Delta_3=5+2n}$ : When $\Delta_3=5+2n$ where $n \in \{0, 1 ,2,..\}$, \eqref{regtriplkconditions} is satisfied for the choice of signs given by $(---)$ and $(++-)$.  Although we have both an ultra-local and a non-local divergence here, the term at $\mathcal{O}(1/\epsilon)$ is non-local in the momenta and therefore the divergence can be cancelled by multiplying with a constant of $\mathcal{O}(\epsilon)$ and then taking the limit $\epsilon \rightarrow 0$. In particular, when $\Delta_3=5$, the divergent term can be calculated to be :
\begin{equation}\label{JJO5}
A_1(k_1, k_2, k_3)=c_1(\epsilon)\frac{k_1+k_2}{k_1 \epsilon}+O(\epsilon^0)
\end{equation}
Choosing $c_1$ to be $O(\epsilon)$, the resulting form factor is :
\begin{equation}\label{jjo5form}
A_1(k_1, k_2, k_3)=c_1^{(1)}\frac{k_1+k_2}{k_1}
\end{equation}
where $c_1^{(1)}$ is $\mathcal O(0)$ in $\epsilon$. 

It can be easily checked that this form factor satisfies non-anomalous Ward identities and that scale invariance is not broken.

\subsection{Non-Abelian $\langle J J O\rangle_{\text{odd}}$}
If we add a colour index to all the three operators in the correlator such that an exchange of two colour indices gives rise to a minus sign, then the symmetry between the form factors in \eqref{JJOansatz2} changes to :
\begin{equation}
\widetilde A(k_1, k_2, k_3)=\widetilde B(k_2, k_1, k_3)
\end{equation}
The solution for $\langle J^{\mu a}(k_1)J^{\nu b}(k_2)O^c_{\Delta}(k_3) \rangle $, when $\Delta=3$ is then given by :
\begin{align}
\langle J^{\mu a}(k_1)J^{\nu b}(k_2)O^c_{\Delta}(k_3) \rangle_{\text{odd}}&=f^{abc}\left[\epsilon^{\nu k_1 k_2}\left(\frac{k_1+k_2+2k_3}{k_1(k_1+k_2+k_3)^2}k_1^{\mu}+\frac{k_1+k_2+2k_3}{k_2(k_1+k_2+k_3)^2}k_2^{\mu}\right)\right.\nonumber\\[7 pt]
&\hspace{.5cm}+\left.\epsilon^{\mu \nu k_2}\left(\frac{k_1+k_2+2k_3}{k_1(k_1+k_2+k_3)^2}k_1^2+\frac{k_1+k_2+2k_3}{k_2(k_1+k_2+k_3)^2}(k_1\cdot k_2)\right)\right]
\end{align}
It can be checked that the above solution is symmetric under $(1 \leftrightarrow 2)$ exchange upon using suitable Schouten identities.

In principle one can compute parity-odd correlation functions of higher spin operators using the approach described in this section following \cite{Bzowski:2013sza}. However, it soon gets difficult to find out the independent tensor structures after the application of the generator of special conformal transformations, due to non-trivial Schouten identities that relate various tensor structures. We will now resort to the technique of using weight-shifting and spin-raising operators to compute parity-odd correlation functions.


\section{Parity-odd spin-raising \& weight-shifting operators in three-dimensions }\label{weight}
In this section we construct parity-odd spin-raising and weight-shifting operators in momentum space. We then illustrate how these operators can be used to calculate the parity-odd part of the $\langle JJO \rangle $ correlator. 

\subsection{Parity-odd operators}
\label{poddoperators}
We consider a parity-odd operator which raises the spins of the operators at points 1 and 2 and lowers the weight of the operator at point 2. In embedding space, this operator is defined as 
\begin{align}
&\widetilde{D}_{12}\equiv\epsilon\left(Z_1, Z_2, X_1, X_2, \frac{\partial}{\partial X_1}\right)
\end{align}
based on requirements of transversality and interiority\cite{Costa:2011dw}. In position space, the operator takes the form :
\begin{align}
\widetilde{D}_{12}= \frac{1}{2}\left(\epsilon^{ijk-+}z_{1i}z_{2j}x_{12k}D_1-\epsilon^{ijk-+}\left[\frac{(x^2_1-x^2_2)}{2}z_{1j}z_{2k}+x_{12j}z_{1k}(z_2\cdot x_2)+z_{2j}x_{12k}(z_1\cdot x_1)\right]P_{1i}\right)
\end{align}
where $\epsilon^{ijk-+} \equiv \epsilon^{ijk}$.
Performing a Fourier transform we get in momentum space :
\begin{align}
&\widetilde{D}_{12}=-\frac{1}{2}\left[\epsilon(z_{1}z_{2}K^-_{12})(\Delta_1-d-k_1\cdot\frac{\partial}{\partial k_1})\notag\right.\\[5pt] &\hspace{2cm}+\left.\frac{K_{12}^-K^+_{12}}{2}\epsilon(k_{1}z_{1}z_{2})+\epsilon(k_{1}K_{12}^-z_{1})(z_2\cdot \frac{\partial}{\partial k_2})+\epsilon(k_{1}z_{2}K^-_{12})(z_1\cdot\frac{\partial}{\partial k_1})\right]\label{FTo}
\end{align}
where $K^+_{12}$ and $K^-_{12}$ are defined in Appendix \ref{ParityEvenOperators}. 

The above operator acts on a momentum space correlator with a momentum conserving delta function. In its present form, it will be tedious to take the above operator past the delta function. Consider now the following commutator : 
\begin{align}
\label{si}
[k^{\mu}_1+k^{\mu}_2+k^{\mu}_3, \widetilde {D}_{12}] &= -\frac{1}{2}[\epsilon(z_1z_2K^{-}_{12})k^{\mu}_1-K^{-\mu}_{12}\epsilon(k_1z_1z_2)-\epsilon(k_1K^{-}_{12}z_1)z^{\mu}_2-\epsilon(k_1z_2K^-_{12})z^{\mu}_1] 
\end{align}
The above commutator vanishes on a three-point function due to momentum conservation. Thus we have the following action on 3-point functions :
\begin{align}
\epsilon(z_1z_2K^{-}_{12})k^{\mu}_1-K^{-\mu}_{12}\epsilon(k_1z_1z_2)-\epsilon(k_1K^{-}_{12}z_1)z^{\mu}_2-\epsilon(k_1z_2K^-_{12})z^{\mu}_1 =0
\end{align}
%
We contract the above equation with $\frac{\partial}{\partial k^{\mu}_1}$ from the right to get the Schouten identity :
\begin{align}
&\epsilon(z_1z_2K^{-}_{12})k_1\cdot\frac{\partial}{\partial k_1}+\epsilon(z_1z_2K^-_{12})-(K^-_{12}\cdot\frac{\partial}{\partial k_1})\epsilon(k_1z_1z_2)-\epsilon(k_1K^{-}_{12}z_1)(z_2\cdot \frac{\partial}{\partial k_1})\notag\\[5pt]&\hspace{2cm}-\epsilon(k_1z_2K^-_{12})(z_1\cdot \frac{\partial}{\partial k_1})=0
\end{align} 
We use this to rewrite \eqref{FTo} as :
\begin{align}\label{D12tilda}
\widetilde{D}_{12}=\epsilon^{z_1 z_2 k_1}W_{12}^{--}+\epsilon^{z_1 k_1 K^-_{12}}(\vec{z}_2\cdot\vec{K}^-_{12})+(2-\Delta_1)\epsilon^{z_1 z_2 K^-_{12}}
\end{align}
where $W_{12}^{--}$ is defined in Appendix \ref{ParityEvenOperators}. Note that the operator defined in \eqref{D12tilda} is explicitly translation invariant\footnote{Translational invariance of an operator implies in momentum space  that its operation on the momentum conserving delta-function is zero. This happens when the operator is only a function of $K^-_{ij}$ in the derivatives, and that is precisely what we have in \eqref{D12tilda}. }.

We will use $\widetilde D_{12}$ in \eqref{D12tilda} to compute $\langle \it{JJ{O}_{\Delta}} \rangle_{\text{odd}}$, $\langle \it{JJJ} \rangle_{\text{odd}}$ and $\langle \it{TT{O}_{\Delta}} \rangle_{\text{odd}}$ \footnote{The subscript on operators denotes their bare scaling dimensions.} starting from a scalar-seed. We can also construct $\widetilde{D}_{23}$ and $\widetilde{D}_{31}$ to get operators that act on points 2 and 3 and points 3 and 1, respectively. We will require them in the computation of $\langle \it{JJJ} \rangle_{\text{odd}}$ as the correlator has cyclic symmetry.


We will now construct other parity-odd weight-shifting and spin-raising operators that are useful. Let us consider the following :
\begin{align}
\widetilde{D}_1&\equiv\epsilon(Z_1, X_1, \frac{\partial}{\partial X^1}, X_2, \frac{\partial}{\partial X_2})+ \epsilon(Z_1, X_1, \frac{\partial}{\partial X^1}, Z_2, \frac{\partial}{\partial Z_2})\notag\\[5pt]&= \frac{1}{2}\big\{\epsilon^{ijk}[z_{1i}x_{12k}(D_2P_{1j}-D_1P_{2j})+(x_1\cdot z_1)x_{12k}P_{1i}P_{2j}-\frac{(x^2_1-x^2_2)}{2}z_{1k}P_{1i}P_{2j}]\big\}\notag\\[5pt]&\hspace{.5cm}+\frac{1}{2}\big\{\epsilon^{ijk}[(x_1\cdot z_1)z_{2j}\frac{\partial}{\partial z^k_2}+(x_2\cdot z_2)z_{1k}\frac{\partial}{\partial z^j_2}+z_{1j}z_{2k}(x_2\cdot\frac{\partial}{\partial z_2})]P_{1i}\big\}\notag\\[5pt]
&\hspace{.5cm}-\frac{1}{2}\epsilon^{ijk}[z_{1i}z_{2j}\frac{\partial}{\partial z^{2k}}]D_1\,.
\end{align}
The Fourier transform of this operator gives in momentum space the following :
\begin{align}
\widetilde{D}_1&=\frac{1}{2}\big\{\epsilon^{ijk}[z_{1i}K^{-}_{12k}((-\Delta_2+d+k_2\cdot\frac{\partial}{\partial k_2})k_{1j}-(-\Delta_1+d+k_1\cdot\frac{\partial}{\partial k_1})k_{2j})\notag\\&\hspace{.5cm}+(z_1\cdot \frac{\partial}{\partial k_1})K^{-}_{12k}k_{1i}k_{2j}-\frac{K^{-}_{12}\cdot K^{+}_{12}}{2}z_{1k}k_{1i}k_{2j}]\big\}\notag\\&\hspace{.5cm}+\frac{1}{2}\big\{\epsilon^{ijk}[(z_1\cdot\frac{\partial}{\partial k_1})z_{2j}\frac{\partial}{\partial z^k_2}+(z_2\cdot\frac{\partial}{\partial k_2})z_{1k}\frac{\partial}{\partial z^j_2}+z_{1j}z_{2k}(\frac{\partial}{\partial k_2}\cdot\frac{\partial}{\partial z_2})]k_{1i}\big\}\notag\\
&\hspace{.5cm}-\frac{1}{2}\epsilon^{ijk}[z_{1i}z_{2j}\frac{\partial}{\partial z^{2k}}](-\Delta_1+d+k_1\cdot\frac{\partial}{\partial k_1})\label{FTD1}
\end{align}
The above can be written in a translation invariant manner using the methods implemented in the case of $\widetilde D_{12}$ to obtain :
\begin{align}
\widetilde{D}_1&=(2-\Delta_2)\epsilon(z_1k_1K^{-}_{12})-(2-\Delta_1)\epsilon(z_1k_2K^{-}_{12})+(z_1\cdot K^{-}_{12})\epsilon(k_1k_2K^{-}_{12})-\epsilon(z_1k_2K^-_{12}) (k_1\cdot K^-_{12})\notag\\[5pt]&\hspace{.5cm}-\epsilon(z_1k_1k_2)\mathcal{W}^{--}_{12}-\epsilon^{ijk}[(z_2\cdot K^{-}_{23})z_{1k}\frac{\partial}{\partial z^j_2}+z_{1j}z_{2k}(\frac{\partial}{\partial z_2}\cdot K^-_{12})]k_{1i}+z_{1i}z_{2j}\frac{\partial}{\partial z^{2k}}(-\Delta_1+d)]
\end{align}
We also introduce the following parity-odd operator :
\begin{align}
D_{123}&=(X_{i}\cdot X_j)\epsilon(Z_1Z_2X_1X_2X_3)\notag\\[5pt]&= \frac{x^2_{ij}}{2}[2(z_2\cdot x_{21})\epsilon(z_1 x_{21} x_{31})+x^2_{12}\epsilon(z_1 x_{31} z_2)+x^2_{31}\epsilon(z_1z_2x_{21})] \quad i, j = 1, 2, 3
\end{align}
which after a Fourier transform takes the form :
\begin{align}
D_{123}=  2[(z_2\cdot K^{-}_{21})\epsilon(z_1 K^{-}_{21} K^{-}_{31})+\epsilon(z_1 K^{-}_{31} z_2)\mathcal{W}^{--}_{12}+\epsilon(z_1z_2K^{-}_{21})\mathcal{W}^{--}_{13}]\mathcal{W}^{--}_{ij}\label{FTD3}
\end{align}
This operator is naturally translational invariant.

Finally, the polarization vectors can be stripped off from the correlators using the Todorov operator \cite{todorov} :
\begin{equation}\label{Todorov}
D_z^{\mu}=\left(\frac{1}{2}+\vec{z}\cdot\frac{\partial}{\partial \vec{z}}\right)\frac{\partial}{\partial z_{\mu}}-\frac{1}{2}z^{\mu}\frac{\partial^2}{\partial \vec{z}\cdot\partial \vec{z}}.
\end{equation}
We will now discuss the computation of $\langle JJO\rangle_{\text{odd}}$ using the technique of spin-raising and weight-shifting operators.
\subsection{Reproducing $\langle JJO\rangle_{\text{odd}}$ correlators }
We start with the seed correlator $\langle O_2 O_3 O_{\Delta} \rangle$, where the last operator has an arbitrary scaling dimension $\Delta$. To get $\langle {JJO_{\Delta}}\rangle_{\text{odd}}$ we act on the scalar-seed with the parity-odd operator $\widetilde{D}_{12}$ defined in \eqref{D12tilda} :
\begin{equation}
\langle J^{\mu}(k_1) J^{\nu}(k_2) O_{\Delta}(k_3) \rangle_{\text{odd}}=D_{z_2}^{\nu}D_{z_1}^{\mu}\widetilde{D}_{12}\langle O_2(k_1)O_3(k_2)O_{\Delta}(k_3) \rangle
\end{equation}
\subsubsection{$\langle J^{\mu}(k_1) J^{\nu}(k_2) O_{2}(k_3) \rangle_{\text{odd}}$}
Let us consider the case when $\Delta=2$. We have :
\begin{equation}
\langle O_2(k_1)O_3(k_2)O_2(k_3) \rangle= c_1\,\text{log}\left(\frac{k_1+k_2+k_3}{\mu}\right)(k_1+k_3)-c_1\,(k_1+k_2+k_3)
\end{equation}
Acting with $\widetilde{D}_{12}$ and then removing the polarization vectors gives
\begin{equation}
\label{jjo2wt}
\langle J^{\mu}(k_1) J^{\nu}(k_2) O_{2}(k_3) \rangle_{\text{odd}}=c_1\frac{\epsilon^{\mu k_1 k_2}(k_1 k_2^\nu-k_2 k_1^\nu)}{k_1\,k_2\,(k_1+k_2+k_3)^2}+c_1\frac{\epsilon^{\mu\nu k_1}(k_1+k_2-k_3)}{2k_1(k_1+k_2+k_3)}
\end{equation}
This matches the answer in \eqref{JJOansatz1}.
To see this explicitly, we substitute the solution to the form factors given in \eqref{JJOdelta2} and \eqref{JJOsec} in the ansatz \eqref{JJOansatz1} :
\begin{equation}
\langle J^{\mu}(k_1) J^{\nu}(k_2) O(k_3) \rangle_{\text{odd}}=c_1\,\frac{\epsilon^{\nu k_1k_2}\,(k_2\,k_1^\mu-k_1\,k_2^\mu)}{k_1k_2(k_1+k_2+k_3)^2}+c_1\,\frac{\epsilon^{\mu\nu k_2}\,(k_1+k_2-k_3)}{2k_2(k_1+k_2+k_3)}
\end{equation}
Using the Schouten identities in \eqref{jjoschouten1} and \eqref{jjoschouten2} to replace $\epsilon^{\nu k_1 k_2}$ in the above equation, we match the result obtained using weight-shifting and spin-raising operators in \eqref{jjo2wt}.
Similarly, one can obtain and match the results for $\langle JJO_1\rangle_{\text{odd}}$ and  $\langle JJO_3\rangle_{\text{odd}}$. 

We were also able to obtain $\langle JJO_{\Delta}\rangle$ for $\Delta = 1, 2, 3$ using the operator in \eqref{FTD1}. However, for $\Delta = 3$ we obtained the correlator up to a conformally invariant contact term given by $\epsilon(z_1z_2k_1)$.

When the dimension of the scalar operator is greater than or equal to 4, one needs to be more careful as the correlators are divergent and need to be renormalised (see Section \eqref{Divergences}). 
 \subsubsection{$\langle J^{\mu}(k_1) J^{\nu}(k_2) O_{4}(k_3)\rangle_{\text{odd}}$}
The full seed correlator along with the divergent part is given by :
\begin{align}\label{JJO4seed}
\begin{split}
\langle {O}_2(k_1){O}_3(k_2){O}_4(k_3) \rangle&=\frac{c_1}{24}\bigg[-11\,k_1^3-6\,k_1^2(k_2+k_3)+3\,k_1(7\,k_2^2+2\,k_2k_3+3\,k_3^2)\bigg.\\[5pt]
&\hspace{.5cm}\bigg.+4\,(4\,k_2^3+3\,k_2^2\,k_3+k_3^3)\bigg.\\[5pt]
&\bigg.\hspace{.5cm}+6\,\log\left(\frac{k_1+k_2+k_3}{\mu}\right)\left(k_1^3-2\,k_2^3-k_1(3\,k_2^2+k_3^2)\right)\bigg]\\[5pt]
&\hspace{.25cm}-c_1\,\frac{1}{\epsilon}\left(k_1^3-2\,k_2^3-k_1(3\,k_2^2+k_3^2)\right)
\end{split}
\end{align}
 This correlator has two semi-local divergences labelled by $(+--)$ and $(-+-)$. The generating functional is given by :
 \begin{equation}
 Z=\int{D\phi\; \text{exp}\left(-\int{d^3x\; \big(S[A_{\mu}, g^{\mu\nu}]+\sqrt{g}(O_2 \phi_1+O_3\phi_0+O_4\phi_{-1})+S_{\text{ct}}\big)}\right)}
 \end{equation}
where $\phi_0$, $\phi_1$ and $\phi_{-1}$ correspond to the sources of $O_3$, $O_2$ and $O_4$ respectively. The counter-term in this case is given by
 \begin{align}
S_{\text{ct}}=\int{d^{3+\epsilon}x\; \mu^{-\epsilon}\,\left(a_1(\epsilon) \Box O_{2} \phi_{0}\phi_{-1}+a_2(\epsilon) O_{2} \Box \phi_{0}\phi_{-1}+a_3(\epsilon)  O_{2} \phi_{0}\Box \phi_{-1}+a_4(\epsilon) O_{3} \phi_{1}\phi_{-1}\right)}
 \end{align}
The full renormalized correlator is then defined by
 \begin{align}
 \begin{split}
 \langle {O}_2(k_1){O}_3(k_2){O}_4(k_3) \rangle_{\text{ren}}&=\frac{-8}{\sqrt{g(x_1)}\sqrt{g(x_2)}\sqrt{g(x_3)}}\frac{\delta}{\delta \phi_1 (x_1)}\frac{\delta}{\delta \phi_0 (x_2)}\frac{\delta}{\delta \phi_{-1} (x_3)}Z\\[7 pt]
 &=c_1\,\frac{1}{24}\Bigg[-11\,k_1^3-6\,k_1^2(k_2+k_3)+3\,k_1(7\,k_2^2+2\,k_2k_3+3\,k_3^2)\bigg.\\
&\bigg.\hspace{1.5cm}+4\,(4\,k_2^3+3\,k_2^2\,k_3+k_3^3)\bigg.\\
&\bigg.\hspace{1.5cm}+6\,\text{log}\left(\frac{k_1+k_2+k_3}{\mu}\right)\left(k_1^3-2\,k_2^3-k_1(3\,k_2^2+k_3^2)\right)\Bigg]
 \end{split}
 \end{align}
We now act $\widetilde{D}_{12}$ from \eqref{D12tilda} on the above renormalised scalar correlator and then strip off the polarization vectors to obtain :
 \begin{equation}
\langle J^{\mu}(k_1) J^{\nu}(k_2) O_{4}(k_3) \rangle_{\text{odd}}=-\epsilon^{k_1 k_2 \mu}k_1^{\nu}\,A_1+\epsilon^{k_1 k_2 \mu}k_2^{\nu}\,A_2+\epsilon^{k_1 \mu\nu}(A_1\,k_1\cdot k_2+A_2\,k_2^2)
 \end{equation}
where $A_1$ is as in \eqref{jjo4form} and $A_2(k_1,k_2,k_3)=-A_1(k_2,k_1,k_3)$. Once again, using the Schouten identities in \eqref{jjoschouten1} and \eqref{jjoschouten2}, this matches the solution \eqref{jjonewansatz} and \eqref{jjo4form} obtained by solving the conformal Ward identities.
\subsubsection{$\langle J^{\mu}(k_1) J^{\nu}(k_2) O_{5}(k_3) \rangle_{\text{odd}}$}
The seed correlator here has ultra-local and non-local divergences. The correlator is given by
\begin{align}\label{jjo5seed}
\langle O_2(k_1)O_3(k_2)O_5(k_3) \rangle&=\frac{c_1}{16\epsilon}\left(5\,k_1^4-40\,k_1\,k_2^3-15\,k_2^4+6\,k_2^2\,k_3^2+k_3^4-6\,k_1^2(5\,k_2^2+k_3^2)\right)+\mathcal{O}(\epsilon^0)
\end{align}
Unlike the previous case, this divergence is removed by taking the constant $c_1$ to be $\mathcal{O}(\epsilon)$ and then taking the limit $\epsilon \rightarrow 0$ (similar to \eqref{JJO5} and \eqref{jjo5form}). Thus we have the renormalised correlator :
\begin{align}
\langle O_2(k_1)O_3(k_2)O_5(k_3) \rangle_{\text{ren}}&=c_1\,\frac{1}{16}\left(5\,k_1^4-40\,k_1\,k_2^3-15\,k_2^4+6\,k_2^2\,k_3^2+k_3^4-6\,k_1^2(5\,k_2^2+k_3^2)\right)
\end{align}
The rest of the calculation is the same as in the case of $\langle JJO_4 \rangle_{\text{odd}}$ and we get 
\begin{equation}
\langle J^{\mu}(k_1) J^{\nu}(k_2) O_{5}(k_3) \rangle_{\text{odd}}=-\epsilon^{k_1 k_2 \mu}k_1^{\nu}\,A_1+\epsilon^{k_1 k_2 \mu}k_2^{\nu}\,A_2+\epsilon^{k_1 \mu\nu}(A_1\,k_1\cdot k_2+A_2\,k_2^2)
\end{equation}
 where $A_1$ is as in \eqref{jjo5form} and $A_2 = -A_1(k_1\leftrightarrow k_2)$. It can be easily checked that this matches the answer obtained by solving conformal Ward identities (see \eqref{jjonewansatz} and \eqref{jjo5form}).
\section{$\langle JJJ\rangle_{\text{odd}}$ in three-dimensions}
\label{JJJsection}
We now turn our attention to computing the odd part of the $\langle JJJ\rangle$ correlator. The correlator is non-zero only when the currents are non-Abelian.

We express $\langle JJJ\rangle_{\text{odd}}$ in terms of transverse and longitudinal parts \cite{Bzowski:2013sza} :
\begin{align}
\label{jjjtransversepluslocal}
\begin{split}
\langle J^{\mu a} J^{\nu b} J^{\rho c} \rangle_{\text{odd}} &= \langle j^{\mu a} j^{\nu b} j^{\rho c} \rangle_{\text{odd}}+\bigg[\frac{k_1^{\mu}}{k_1^2}\left(f^{adc}\langle J^{\rho d}(k_2)J^{\nu b}(-k_2) \rangle-f^{abd}\langle J^{\nu d}(k_3)J^{\rho c}(-k_3)\rangle\right)\bigg.\\[5 pt]
&\bigg.+\left((k_1, \mu) \leftrightarrow (k_2, \nu)\right)+\left((k_1, \mu) \leftrightarrow (k_3, \rho)\right)\bigg]+\bigg[\left(\frac{k_1^{\mu}\,k_2^{\nu}}{k_1^2 k_2^2}f^{abd}k_{2\alpha}\langle J^{\alpha d}(k_3)J^{\rho c}(-k_3)\rangle\right)\bigg.\\[5 pt]
&\bigg.+\left((k_1, \mu) \leftrightarrow (k_3, \rho)\right)+\left((k_2, \nu) \leftrightarrow (k_3, \rho)\right)\bigg]
\end{split}
\end{align}
where $ \langle j^{\mu a} j^{\nu b} j^{\rho c} \rangle_{\text{odd}}$ denotes the transverse part of the correlator. The ansatz for this part of the correlator can be written as
\begin{equation}
 \langle j^{\mu a} j^{\nu b} j^{\rho c} \rangle_{\text{odd}}=\pi^{\mu}_{\alpha}(k_1)\pi^{\nu}_{\beta}(k_2)\pi^{\rho}_{\gamma}(k_3)X^{\alpha\beta\gamma}
\end{equation}
where
\begin{align}\label{jjjansatz}
\begin{split}
X^{\alpha\beta\gamma}&=A_1\epsilon^{k_1 k_2 \alpha}k_1^{\gamma}k_3^{\beta}+A_2\epsilon^{k_1 k_2 \alpha}\delta^{\beta\gamma}+A_3\epsilon^{k_1 \alpha \beta}k_1^{\gamma}+A_4\epsilon^{k_1 \alpha \gamma}k_3^{\beta}+\text{cyclic perm.}
\end{split}
\end{align}
Calculating the form factors by directly solving the conformal Ward identities is quite complicated. Here we will instead use spin-raising and weight shifting operators to calculate them.

Starting from the seed correlator $\langle O_3(k_1)O_2(k_2)O_3(k_3) \rangle$ we can get $\langle JJJ\rangle_{\text{odd}}$ by :
\begin{align}\label{jjjfromseed}
\langle J^{\mu a}(k_1)J^{\nu b}(k_2)J^{\rho c}(k_3) \rangle_{\text{odd}}=\frac{1}{4}D^{\mu}_{z_1}D^{\nu}_{z_2}D^{\rho}_{z_3}\widetilde{D}_{23}D_{11}\langle O^a_3(k_1)O^b_2(k_2)O^c_3(k_3) \rangle_{\text{even}}+\text{cyclic perm.}
\end{align}
where $\widetilde{D}_{12}$ is defined in \eqref{D12tilda} and $D_{11}$ in \eqref{D11}.
The renormalized seed correlator is :
\begin{align}\label{jjjseed}
\langle O^a_3(k_1)O^b_2(k_2)O^c_3(k_3) \rangle=&f^{abc}c_1\bigg[2\,\text{log}\left(\frac{k_1+k_2+k_3}{\mu}\right)(k_1^2-k_2^2+k_3^2)-k_1^2-k_3^2\bigg.\nonumber\\[6 pt]
&\bigg.\hspace{1.5cm}+2\,k_1(k_2-k_3)+2\,k_2k_3+3\,k_2^2\bigg]
\end{align}
This gives :
\begin{align}\label{jjjanswer}
\begin{split}
X^{\alpha\beta\gamma}=&-\frac{2}{k_1(k_1+k_2+k_3)^3}\epsilon^{k_1 k_2 \alpha}k_1^{\gamma}k_3^{\beta}-\frac{1}{(k_1+k_2+k_3)^2}\epsilon^{k_1 k_2 \alpha}\delta^{\beta\gamma}\\[5 pt]
&+\frac{(k_1+k_2+2\,k_3)}{k_1(k_1+k_2+k_3)^2}\epsilon^{k_1 \alpha\beta}k_1^{\gamma}+\frac{(k_1+2\,k_2+k_3)}{k_1(k_1+k_2+k_3)^2}\epsilon^{k_1 \alpha\gamma}k_3^{\beta}\\[5 pt]
&+\text{cyclic perm.}
\end{split}
\end{align}
We can now read off the form factors in \eqref{jjjansatz} by comparing it with \eqref{jjjanswer} :
\begin{align}
A_1&=-\frac{2}{k_1(k_1+k_2+k_3)^3},\quad\quad
A_2=-\frac{1}{(k_1+k_2+k_3)^2}\nonumber\\[5 pt]
A_3&=\frac{k_1+k_2+2\,k_3}{k_1(k_1+k_2+k_3)^2},\quad\quad\,\,\,\,
A_4=\frac{k_1+2\,k_2+k_3}{k_1(k_1+k_2+k_3)^2}
\end{align}
The $\langle \it{JJJ} \rangle_{\text{odd}}$ correlator obeys the following Ward-Takahashi identity \cite{Bzowski:2013sza} :
\begin{align}\label{jjjWT}
k_{1\mu}\langle J^{\mu a}(k_1)J^{\nu b}(k_2)J^{\rho c}(k_3) \rangle_{\text{odd}}&=f^{adc}\langle J^{\rho d}(k_2)J^{\nu b}(-k_2)\rangle_{\text{odd}}-f^{abd}\langle J^{\nu d}(k_3)J^{\rho c}(-k_3)\rangle_{\text{odd}}\nonumber\\[5pt]
&=-f^{abc}\left(\epsilon^{k_2 \nu \rho}+\epsilon^{k_3 \nu \rho}\right)
\end{align}
Our result does satisfy this identity. To see this, let us contract our result  \eqref{jjjtransversepluslocal} for $\langle J^{\mu a}J^{\nu b}J^{\rho c}\rangle$ with $k_1^{\mu}$ :
\begin{align}
\begin{split}
k_{1\mu}\langle J^{\mu a}J^{\nu b}J^{\rho c}\rangle_{\text{odd}}&= f^{abc}\,\epsilon^{k_1 k_2 \nu}(B_1 k_1^{\rho}+(B_1-B_1(k_2 \leftrightarrow k_3))k_3^{\rho})\\[5 pt]
&\hspace{.5cm}+f^{abc}\,\epsilon^{k_1 k_2 \rho}(B_1 k_2^{\nu}+B_1(k_2 \leftrightarrow k_3) k_3^{\nu})-2\left(\epsilon^{k_2 \nu\rho}B_2-\epsilon^{k_3 \nu\rho}B_2(k_2 \leftrightarrow k_3)\right)
\end{split}
\end{align}
where
\begin{align}
\begin{split}
B_1=\frac{4(2k_1+k_2+k_3)}{k_2(k_1+k_2+k_3)^2},\quad
B_2=\frac{2k_1^2+k_2^2+k_1(k_2-k_3)-k_3^2}{k_2(k_1+k_2+k_3)}
\end{split}
\end{align}
Using the identities in \eqref{jjjschouten1} and \eqref{jjjschouten2}, we can get rid of those epsilon structures which have two momenta contracted with their indices. Doing so we obtain : 
\begin{equation}
k_{1\mu}\langle J^{\mu a}J^{\nu b}J^{\rho c}\rangle=-f^{abc}(\epsilon^{k_2 \nu\rho}+\epsilon^{k_3 \nu\rho})
\end{equation}
which matches the desired Ward identity \eqref{jjjWT}.

\section{$\langle TTO\rangle_{\text{odd}}$ in three-dimensions}\label{TTO}
We now turn our attention to the $\langle TTO\rangle_{\text{odd}}$ correlator. Based on symmetry considerations and conservation, this correlator is expected to take the following form \cite{Bzowski:2013sza} :
\begin{align}
&\langle T^{\mu_{1} \nu_{1}}(k_1) T^{\mu_{2} \nu_{2}}(k_2) {O}^{I}(k_3)\rangle_{\text{odd}}=\langle t^{\mu_{1} \nu_{1}}(k_1) t^{\mu_{2} \nu_{2}}(k_2) {O}^{I}(k_3)\rangle_{\text{odd}}\notag\\[5pt]
&\quad+2\left[\mathcal{T}^{\mu_{1} \nu_{1} \alpha_{1}}(k_1) k_{1}^{\beta_{1}}+\frac{\pi^{\mu_{1} \nu_{1}}(k_1)}{d-1} \delta^{\alpha_{1} \beta_{1}}\right] \delta^{\mu_{2} \alpha_{2}} \delta^{\nu_{2} \beta_{2}}\left\langle\frac{\delta T_{\alpha_1 \beta_1}}{\delta g^{\alpha_{2} \beta_{2}}}(k_1, k_2){O}^{I}(k_3)\right\rangle\notag\\[5pt]&\quad+2\left[\left(\mu_{1}, \nu_{1}, k_1\right) \leftrightarrow\left(\mu_{2}, \nu_{2},k_2 \right)\right]\notag\\[5pt]&
~~~-4\left[\mathcal{T}^{\mu_{1} \nu_{1} \alpha_{1}}(k_1) k_{1}^{\beta_{1}}+\frac{\pi^{\mu_{1} \nu_{1}}(k_1)}{d-1} \delta^{\alpha_{1} \beta_{1}}\right]\left[\mathcal{T}^{\mu_{2} \nu_{2} \alpha_{2}}(k_2) k_{2}^{\beta_{2}}+\frac{\pi^{\mu_{2} \nu_{2}}(k_2)}{d-1} \delta^{\alpha_{2} \beta_{2}}\right]\notag\\[5pt]&~~~\times \left\langle\frac{\delta T_{\alpha_1 \beta_1}}{\delta g^{\alpha_{2} \beta_{2}}}(k_1, k_2) {O}^{I}(k_3)\right\rangle\label{rf}
\end{align}
where
\begin{align}
\mathcal{T}^{\mu\nu\alpha}(p) = \eta^{\alpha\beta}\mathcal{T}_{\beta}^{\mu\nu}= \frac{ \eta^{\alpha\beta}}{p^{2}}\left[2 p^{(\mu} \delta_{\beta}^{\nu)}-\frac{p_{\beta}}{d-1}\left(\delta^{\mu \nu}+(d-2) \frac{p^{\mu} p^{\nu}}{p^{2}}\right)\right]
\end{align}
and 
 $\langle t^{\mu_{1} \nu_{1}}(k_1) t^{\mu_{2} \nu_{2}}(k_2) {O}^{I}(k_3)\rangle_{\text{odd}}$ is the transverse part of the correlator. Due to symmetry and transversality, this is expected to take the following form :
\begin{align}
\label{ttoform}
&\langle t^{\mu_{1} \nu_{1}}(k_1) t^{\mu_{2} \nu_{2}}(k_2) {O}^{I}(k_3)\rangle_{\text{odd}}=\notag\\[5pt]&\Pi^{\mu_1\nu_1}_{\alpha_1\beta_1}(k_1)\Pi^{\mu_2\nu_2}_{\alpha_2\beta_2}(k_2)\left[A_{11}\epsilon^{\alpha_1 k_1 k_2}k_2^{\beta_1}k_1^{\beta_2}k_1^{\alpha_2}+A_{12}\epsilon^{\alpha_2 k_1 k_2}k_2^{\beta_1}k_1^{\beta_2}k_2^{\alpha_1}+A_{21}\epsilon^{\alpha_1 k_1 k_2}\delta^{\beta_1\beta_2}k_1^{\alpha_2}\right.\notag\\[5 pt]
&\hspace{3.5cm}\left.+A_{22}\epsilon^{\alpha_2 k_1 k_2}\delta^{\beta_1\beta_2}k_2^{\alpha_1}+A_{31}\epsilon^{\alpha_1 \alpha_2 k_1}k_2^{\beta_1}k_1^{\beta_2}+A_{32}\epsilon^{\alpha_1 \alpha_2 k_2}k_2^{\beta_1}k_1^{\beta_2}\right.\notag\\[5 pt]
&\hspace{3.5cm}\left.+A_{41}\epsilon^{\alpha_1\alpha_2 k_1}\delta^{\beta_1\beta_2}+A_{42}\epsilon^{\alpha_1\alpha_2 k_2}\delta^{\beta_1\beta_2}\right].
\end{align}
where the traceless-orthogonal projector is given by :
\begin{align}
	\Pi_{\alpha \beta}^{\mu \nu}({p})&\equiv\frac{1}{2}\left(\pi_{\alpha}^{\mu}({p})\,\pi_{\beta}^{\nu}({p})+\pi_{\beta}^{\mu}({p})\,\pi_{\alpha}^{\nu}({p})\right)-\frac{1}{2} \pi^{\mu\nu}({p})\,\pi_{\alpha \beta}({p})\,.\label{projk1k1}
	\end{align}
In the following, we first present a direct calculation of $\langle TTO\rangle$ in the free-fermion theory for which $O={\bar \psi}\psi$ which is parity-odd in three-dimensions and has scaling dimension $\Delta=2$. We then reproduce the free theory answer using spin and weight-shifting operators. We also compute $\langle TTO\rangle_{\text{odd}}$  with $\Delta_{O}=1$ which does not have a free theory analogue \footnote{One can get such a parity-odd correlator by coupling a complex scalar field theory to a Chern-Simons gauge field. See \cite{Aharony:2011jz,Giombi:2011kc,Maldacena:2012sf,Aharony:2012nh,GurAri:2012is}
}. We defer the computation of $\langle TTO_{\Delta}\rangle $ for general $\Delta$ to future work.
\subsection{$\langle TTO_2\rangle$ for free fermionic theory}
\label{TTO2FFsection}
The free-fermion action in curved space is given by \cite{Bzowski:2013sza} :
\begin{align}
\label{FFaction}
S= \int d^3x ~e \left[ \bar{\psi} e^{\mu}_{a}\gamma^a \stackrel{\leftrightarrow}{\nabla}_{\mu}\psi \right]
\end{align}
where $e^a_{\mu}$ are the vielbeins and the covariant derivative acts as follows on the spinor
\begin{align}
&\nabla_{\mu}\psi = (\partial_{\mu}-\frac{i}{2}\omega^{ab}_{\mu}\Sigma_{ab})\psi\notag\\[5pt]
&\bar{\psi}\stackrel{\leftarrow}{\nabla}_{\mu} = \bar{\psi} (\stackrel{\leftarrow}{\partial}_{\mu} +\frac{i}{2}\omega^{ab}_{\mu}\Sigma_{ab})\label{cov}
\end{align}
One may use this action to compute the stress-energy tensor \cite{Bzowski:2013sza} :
\begin{align}
\label{FFSET}
T_{\mu \nu}=\frac{1}{\sqrt{g}} \frac{\delta S}{\delta g^{\mu \nu}}=\frac{1}{2}\bar{\psi} \gamma_{(\mu} \stackrel{\leftrightarrow}{\nabla}_{\nu)} \psi
\end{align}
which after taking the flat-space limit and a Fourier transform gives the stress-energy tensor in momentum space :
\begin{align}
\label{op}
T_{\mu\nu}(k) = \frac{1}{4}\int d^3 l\bar{\psi}(l)[\gamma_{\mu}(2l-k)_{\nu}+\gamma_{\nu}(2l-k)_{\mu}]\psi(k-l)
\end{align}
The parity-odd scalar primary in the free-fermion theory is given by :
\begin{align}
\label{op11}
O_2=\bar{\psi}\psi
\end{align}
Using these definitions it is straightforward to evaluate $\langle TTO_2\rangle$ for the free-fermionic theory. Since $O_2$ is parity-odd $\langle TTO_2\rangle$ is also parity-odd. We give the details of the computation in Appendix \ref{CD} and give only the final results here.

The form factors that appear in the transverse part of the correlator \eqref{ttoform} are :
\begin{align}
A_{11}&=\frac{k_1+4\,k_2+k_3}{6(k_1+k_2+k_3)^4}\nonumber\\[5 pt]
A_{21}&=\frac{2\,k_1^2+4\,k_2^2+3\,k_2k_3+k_3^2+3\,k_1(2\,k_2+k_3)}{6(k_1+k_2+k_3)^3}\nonumber\\[5 pt]
A_{31}&=\frac{k_3(k_1+3\,k_2+k_3)}{4(k_1+k_2+k_3)^3}\nonumber\\[5 pt]
A_{41}&=\frac{k_1^3+2\,k_1^2(k_2+k_3)+2\,k_3^2(2\,k_2+k_3)+k_1(k_2^2+2\,k_2k_3+3\,k_3^2)}{8(k_1+k_2+k_3)^2}\nonumber\\[5 pt]
A_{i2}&=-A_{i1}(k_1 \leftrightarrow k_2).
\end{align}
To compute the longitudinal part, we require the functional derivative of the fermionic stress-energy tensor. 
Using the action of the covariant derivative on spinors \eqref{cov} we re-express the stress-energy tensor \eqref{FFSET} as :
\begin{align}
T_{\mu \nu} =\frac{1}{2} \bar{\psi} \gamma_{(\mu} \stackrel{\leftrightarrow}{\partial}_{\nu)} \psi +\frac{1}{16}\omega_{(\mu}^{~ab}\bar{\psi}\{\gamma_{\nu)},\gamma_{ab}\}\psi
\end{align}
After some computation (see Appendix \ref{TTO2long}) we obtain :
\begin{align}
\label{TTO2int}
\frac{\delta T_{\mu\nu}(x)}{\delta g_{\alpha\beta}(y)} =-\frac{i}{16}\{[\epsilon_{\sigma\alpha\mu}\delta_{\beta\nu}+\epsilon_{\sigma\beta\mu}\delta_{\alpha\nu}+\epsilon_{\sigma\alpha\nu}\delta_{\beta\mu}+\epsilon_{\sigma\beta\nu}\delta_{\alpha\mu}]\partial^{\sigma}\delta^{(3)}(x-y)\}O_2(x)
\end{align}
Taking a Fourier transform of the above we get :
\begin{align}
\frac{\delta T_{\mu\nu}}{\delta g_{\alpha\beta}}(k_1, k_2) =\frac{1}{32}[\epsilon_{k_2\alpha\mu}\delta_{\beta\nu}+\epsilon_{k_2\beta\mu}\delta_{\alpha\nu}+\epsilon_{k_2\alpha\nu}\delta_{\beta\mu}+\epsilon_{k_2\beta\nu}\delta_{\alpha\mu}]O_2(k_3)-(k_1 \leftrightarrow k_2)
\end{align}
Using the above, one may compute
\begin{align}
\left\langle \frac{\delta T_{\mu\nu}}{\delta g_{\alpha\beta}}(k_1, k_2) O_2(-k_3)\right\rangle
=-\frac{k_3}{256}[\epsilon_{k_2\alpha\mu}\delta_{\beta\nu}+\epsilon_{k_2\beta\mu}\delta_{\alpha\nu}+\epsilon_{k_2\alpha\nu}\delta_{\beta\mu}+\epsilon_{k_2\beta\nu}\delta_{\alpha\mu}-(k_1\leftrightarrow k_2)] \label{TWI}
\end{align}
where we used $\langle O_2(k_3)O_2(-k_3)\rangle = -\frac{k_3}{8}$ in the free-fermion theory. This can now be used in the reconstruction formula \ref{rf} to get the full correlator $\langle TTO_2\rangle$.

Having computed the functional derivative, we now give the trace and transverse Ward identities associated with $\langle TTO_2\rangle$. They are \cite{Bzowski:2013sza} :
\begin{align}
&\langle T(k_1)T_{\alpha\beta}(k_2)O_2(k_3)\rangle = 2\left\langle\frac{\delta T}{\delta g^{\alpha\beta}}(k_1, k_2)O(k_3)\right\rangle \label{TrWI}\\[5pt]
&k^{\mu}_1\langle T_{\mu\nu}(k_1)T_{\alpha\beta}(k_2)O_2(k_3)\rangle = 2k^{\mu}_1\left\langle\frac{\delta T_{\mu\nu}}{\delta g^{\alpha\beta}}(k_1, k_2)O(k_3)\label{TaWI}\right\rangle
\end{align}
The expression obtained in \eqref{TWI} is traceless in $(\mu, \nu)$ and $(\alpha, \beta)$. This immediately implies that the trace Ward identity is trivial :
\begin{align}
\langle T(k_1)T_{\alpha\beta}(k_2)O(k_3)\rangle = 0\label{TrWIE}
\end{align}
Contracting \eqref{TWI} with $k^{\mu}_{1}$ gives the transverse Ward identity :
\begin{align}
\label{fdWI}
k^{\mu}_1\langle T_{\mu\nu}(k_1)T_{\alpha\beta}(k_2)O_2(k_3)\rangle =\frac{k_3}{128}\left[-\epsilon_{k_1k_2\alpha}\delta_{\beta\nu}-\epsilon_{k_1k_2\beta}\delta_{\alpha\nu}+k_{1\beta}(\epsilon_{k_1\alpha\nu}-\epsilon_{k_2\alpha\nu})+k_{1\alpha}(\epsilon_{k_1\beta\nu}-\epsilon_{k_2\beta\nu})\right]
\end{align}
%
%
%
In  Appendix \eqref{WI} we show that precisely this Ward identity holds.
\subsection{$\langle TTO_2\rangle_{\text{odd}}$ using parity-odd spin-raising and weight-shifting operators}
In this section we compute the odd part of $\langle TTO_2\rangle$ using spin-raising and weight-shifting operators.
We start from the renormalised scalar-seed correlator $\langle O_4(k_1)O_5(k_2)O_2(k_3)\rangle$ given by :
\begin{align}\label{o4o5o2seed}
\langle O_4(k_1)O_5(k_2)O_2(k_3)\rangle&=\frac{1}{960}\Big[-736k_1^5-60k_1^3(7k_2-9k_3)k_3-180k_1(k_2-k_3)^2k_3(k_2+k_3)\nonumber\\[5pt]
&\hspace{.5cm}-15k_1^4(32k_2+77k_3)-10k_1^2(16k_2^3+57k_2^2k_3-42k_2k_3^2-119k_3^3)\nonumber\\[5pt]
&\hspace{.5cm}-3(k_2+k_3)^2(32k_2^3+41k_2^2k_3-214k_2k_3^2+137k_3^3)\nonumber\\[5pt]
&\hspace{.5cm}+60\log\left(\frac{k_1+k_2+k_3}{\mu}\right)\left(8k_1^5+15k_1^4k_3+3k_3(k_2^2-k_3^2)^2+2k_1^2(3k_2^2k_3-5k_3^3)\right)\Big]
\end{align}
%
%
The following sequence of operations gives the  $\langle TTO_2\rangle$ correlator :
\begin{align}
\label{TTO2sequence}
&\widetilde D_{12}(a_1\,D_{22}D_{11}+a_2\,D_{21}D_{12}+a_3\,H_{12}+a_4\,W_{12}^{--}S_{12}^{++})\langle O_4(k_1)O_5(k_2)O_2(k_3)\rangle\nonumber\\[5pt]
&\hspace{.5cm}+(b_1\,D_{22}D_{11}+b_2\,D_{21}D_{12}+b_3\,H_{12}+b_4\,W_{12}^{--}S_{12}^{++})\widetilde D_{12}\langle O_4(k_1)O_5(k_2)O_2(k_3)\rangle\nonumber\\[5pt]
&\hspace{.5cm}+(k_1\leftrightarrow k_2,z_1\leftrightarrow z_2)
\end{align}
where $a_i$ and $b_i$ are coefficients which are fixed by comparing with the result for the correlator from the explicit computation in the free fermion theory :
\begin{align}
a_1&=-\frac{1}{45},\quad
a_2=-\frac{1}{45}+a_4,\quad
a_3=-\frac{2}{45}+8(a_4+b_4),\nonumber\\[5pt]
b_1&=0,\quad\quad
b_2=b_4+\frac{1}{90},\quad
b_3=2b_4+\frac{1}{60}
\end{align}
It can be checked that the operators in \eqref{TTO2sequence} are not all independent. One has the following linear dependence between them :
\begin{align}
\widetilde D_{12}(D_{21}D_{12}+8\,H_{12}+W_{12}^{--}S_{12}^{++})\langle O_4(k_1)O_5(k_2)O_2(k_3)\rangle&=0\nonumber\\[5pt]
(D_{21}D_{12}+10\,H_{12}+W_{12}^{--}S_{12}^{++})\widetilde D_{12}\langle O_4(k_1)O_5(k_2)O_2(k_3)\rangle&=0
\end{align}
Using these relations and setting $b_4=\frac{1}{180}$ one obtains the odd part of $\langle TTO_2\rangle$ correlator from the following :
\begin{align}\label{tto2sequence}
\langle TTO_2\rangle_{\text{odd}}=\left[-4\widetilde D_{12}(D_{22}D_{11}+D_{21}D_{12})+\left(2 D_{21}D_{12}-5H_{12}\right)\widetilde D_{12}\right]\langle O_4(k_1)O_5(k_2)O_2(k_3)\rangle
\end{align}

\subsection{$\langle TTO_1\rangle_{\text{odd}}$ using parity-odd spin-raising and weight-shifting operators}
In this subsection we discuss the parity-odd part of the correlator $\langle TTO_1\rangle$. To construct this correlator we start from the renormalised scalar seed correlator $\langle O_4(k_1)O_5(k_2)O_1(k_3)\rangle$ given by :
\begin{align}
\langle O_4(k_1)O_5(k_2)O_1(k_3)\rangle&=\frac{1}{960k_3}\Big[-736k_1^5-60k_1^3(7k_2-9k_3)k_3-180k_1(k_2-k_3)^2k_3(k_2+k_3)\nonumber\\[5pt]
&\hspace{.5cm}-15k_1^4(32k_2+77k_3)-10k_1^2(16k_2^3+57k_2^2k_3-42k_2k_3^2-119k_3^3)\nonumber\\[5pt]
&\hspace{.5cm}-3(k_2+k_3)^2(32k_2^3+41k_2^2k_3-214k_2k_3^2+137k_3^3)\nonumber\\[5pt]
&\hspace{.5cm}+60\log\left(\frac{k_1+k_2+k_3}{\mu}\right)\left(8k_1^5+15k_1^4k_3+3k_3(k_2^2-k_3^2)^2+2k_1^2(3k_2^2k_3-5k_3^3)\right)\Big]
\end{align}
Note that this is related to the scalar-seed $\langle O_4(k_1)O_5(k_2)O_1(k_3)\rangle$ in \eqref{o4o5o2seed} which was used  to compute $\langle TTO_2\rangle$ as :
\begin{align}
\label{o4o5o1vso4o5o2}
\langle O_4(k_1)O_5(k_2)O_1(k_3)\rangle=\frac{1}{k_3}\langle O_4(k_1)O_5(k_2)O_2(k_3)\rangle
\end{align}
%
The correlator $\langle TTO_1\rangle$ is given by the same sequence of operations \eqref{tto2sequence} acting on $\langle O_4(k_1)O_5(k_2)O_1(k_3)\rangle$
\begin{align}
\langle TTO_1\rangle_{\text{odd}}=\left[-4\widetilde D_{12}(D_{22}D_{11}+D_{21}D_{12})+\left(2 D_{21}D_{12}-5 H_{12}\right)\widetilde D_{12}\right]\langle O_4(k_1)O_5(k_2)O_1(k_3)\rangle
\end{align}
Note that the spin-raising and weight-shifting operators that appear in the above equation are independent of the third momentum and hence do not act on the $k_3$ pre-factor in \eqref{o4o5o1vso4o5o2}. As a result we obtain the following relation
\footnote{This is consistent with the fact that the two scalar operators with $\Delta=1$ and $\Delta=2$ are related by a shadow transformation in three-dimensions.}:
\begin{align}\label{TO2O1}
\langle TTO_1\rangle_{\text{odd}}=\frac{1}{k_3}\langle TTO_2\rangle_{\text{odd}}
\end{align}



\section{Momentum space parity-odd 3-point correlators in 4d}
\label{4dsection}
In this section we will construct and use parity-odd spin-raising operators in four-dimensions to determine the non-trivial momentum-space correlator $\langle JJJ \rangle $.  In four-dimensions, the only non-zero parity-odd 3-point function is $\langle JJJ\rangle $ with a nonabelian current $J.$ Other correlators involving operators with spin $s\le 2$ are zero as emphasized in subsection \ref{4dpodd}.
\subsection{Spin-raising and weight-shifting operators}
In four-dimensions, transversality and interiority allow for only one operator :
\begin{align}
D^{++}_{12}&\equiv\epsilon(Z_1Z_2X_1X_2\frac{\partial}{\partial X_1}\frac{\partial}{\partial X_2})\notag\\[5pt]&=\frac{1}{2}\big\{-\epsilon^{ijkl}z_{1j}z_{2k}x_{12l}(D_1P_{2i}-D_2P_{1i})\notag\\[5pt]&\hspace{.5cm}+\epsilon^{ijkl}[(z_2\cdot x_2)z_{1k}x_{12l}-(z_1\cdot x_1)z_{2k}x_{12l}-\left(\frac{x^2_1-x^2_2}{2}\right)z_{1k}z_{2l}]P_{1i}P_{2j}\big\}
\end{align}
where the embedding space result has been converted to ordinary position space. Fourier transform of the above operator gives :
\begin{align}
&D^{++}_{12}=\frac{1}{2}\big\{-\epsilon^{ijkl}z_{1j}z_{2k}K^{-}_{12l}[(-\Delta_1+d+k_1\cdot\frac{\partial}{\partial k_1})k_{2i}-(-\Delta_2+d+k_2\cdot\frac{\partial}{\partial k_2})k_{1i}]\notag\\[5pt]&+\epsilon^{ijkl}[(z_2\cdot\frac{\partial}{\partial k_2})z_{1k}K^{-}_{12l}-(z_1.\frac{\partial}{\partial k_1})z_{2k}K^{-}_{12l}-\frac{K^-_{12}K^+_{12}}{2}z_{1k}z_{2l}]k_{1i}k_{2j}\big\}
\end{align}
Let us now consider the following commutator :
\begin{align}
&[k^{\mu}_1+k^{\mu}_2+k^{\mu}_3, D^{++}_{12}] =\notag\\[5pt]&-\epsilon(z_1z_2K^-_{12}k_2)k^{\mu}_1+\epsilon(z_1z_2K^-_{12}k_1)k^{\mu}_2-z^{\mu}_{2}\epsilon(z_1k_1k_2K^-_{12})+z^{\mu}_{1}\epsilon(z_2k_1k_2K^-_{12})+K^{-\mu}_{12}\epsilon(z_1z_2k_1k_2)
\end{align}
which is zero by momentum conservation. The above commutator gives an operator-based Schouten identity. We find the following contracted form of the above equation useful in the present context :
\begin{align}
&-\epsilon(z_1z_2K^-_{12}k_2)k_1\cdot\frac{\partial}{\partial k_1}+\epsilon(z_1z_2K^-_{12}k_1)k_2\cdot\frac{\partial}{\partial k_1}-3\epsilon(z_1z_2K^-_{12}k_2)\notag\\[5pt]&\hspace{1cm}-z_{2}\cdot\frac{\partial}{\partial k_1}\epsilon(z_1k_1k_2K^-_{12})+z_{1}\cdot\frac{\partial}{\partial k_1}\epsilon(z_2k_1k_2K^-_{12})+K^{-}_{12}\cdot\frac{\partial}{\partial k_1}\epsilon(z_1z_2k_1k_2)=0
\end{align}
The above contracted form allows us to write a manifestly translation invariant form of the spin-raising operator :
\begin{align}
\label{Dpp12}
D^{++}_{12}&=\frac{1}{2}\big\{\epsilon(z_{1}z_{2}K^{-}_{12}k_2)(-\Delta_1+2)-\epsilon(z_{1}z_{2}K^{-}_{12}k_1)(-\Delta_2+3)-(z_2.K^-_{12})\epsilon(z_{1}k_1k_2K^-_{12})\notag\\[5pt]&\hspace{1cm}+\epsilon(z_1z_2K^-_{12}k_1)(k_2\cdot K^-_{12})+\epsilon(z_{1}z_{2}k_{1}k_{2})\mathcal{W}^{--}_{12}\big\}
\end{align}
As in \eqref{D12tilda} the operator depends only on $K^-_{ij}$ in the derivatives, ensuring translational invariance.
\subsection{$\langle JJJ\rangle_{\text{odd}}$}
In this section we use the operator in \eqref{Dpp12} to derive the odd part of $\langle JJJ\rangle$. In four-dimensions, this correlator is expected to have the following structure based on transversality and momentum conservation :
\begin{align}
\langle J^a_{\mu}(k_1)J^b_{\nu}(k_2)J^c_{\lambda}(k_3)\rangle_{\text{odd}} &= \pi_{\mu}^{\alpha}(k_1)\pi_{\nu}^{\beta}(k_2)\pi_{\lambda}^{\gamma}(k_3)[A^{abc}\epsilon_{\alpha\beta\gamma k_1}+B^{abc}\epsilon_{\alpha\beta\gamma k_2}\notag\\[5pt]&+C^{abc} k_{1\gamma}\epsilon_{\alpha\beta k_1 k_2}+D^{abc}k_{2\alpha}\epsilon_{\beta\gamma k_1 k_2}+E^{abc}k_{1\beta}\epsilon_{\gamma\alpha k_1 k_2} ]
\end{align}
The above ansatz can be simplified using the Schouten identities in \eqref{schouten4D} to the following
\begin{align}
\label{jjj4dnonabelian}
\langle J^a_{\mu}(k_1)J^b_{\nu}(k_2)J^c_{\lambda}(k_3)\rangle_{\text{odd}}= \pi_{\mu}^{\alpha}(k_1)\pi_{\nu}^{\beta}(k_2)\pi_{\lambda}^{\gamma}(k_3)\left[C^{abc} k_{1\gamma}\epsilon_{\alpha\beta k_1 k_2}+D^{abc}k_{2\alpha}\epsilon_{\beta\gamma k_1 k_2}+E^{abc}k_{1\beta}\epsilon_{\gamma\alpha k_1 k_2}\right]
\end{align}
Due to cyclic symmetry we obtain the following relations between the form factors
\begin{align}
&C^{abc}(k_1, k_2, k_3) = -E^{bca}( k_2, k_3, k_1)\nonumber\\[5pt]& E^{abc}(k_1, k_2, k_3) = -D^{bca}(k_2, k_3, k_1)\nonumber\\[5pt]& D^{abc}(k_1, k_2, k_3) = -C^{bca}(k_2, k_3, k_1) 
\end{align}
This shows that there is only one independent form factor. To compute $ \langle JJJ\rangle_{\text{odd}}$ in four-dimensions one has to compute :
\begin{align}
 \langle JJJ\rangle_{\text{odd}}=D^{++}_{12}\langle O_3O_3 J\rangle+\text{cyclic perms.}
\end{align}
The correlator $\langle O_3O_3 J\rangle$ has the following ansatz :
\begin{align}
\label{reco}
\langle O(k_1) O(k_2) J_{\mu}(k_3)\rangle = \langle O(k_1)O(k_2)j_{\mu}(k_3)\rangle-\frac{k_{3\mu}}{k^2_3}(\langle O(k_1)O(-k_1)\rangle+\langle O(k_2)O(-k_2)\rangle)
\end{align}
where $\langle OOj\rangle$ is the transverse part, which has the following ansatz
\begin{align}
\langle O(k_1)O(k_2)j_{\mu}(k_3)\rangle = A(k_1, k_2, k_3)k^{\nu}_2\pi_{\mu\nu}(k_3)
\end{align}
The form factor $A$ is computed by solving the conformal ward identities, and is given by
\begin{align}
A(k_1, k_2, k_3) = 2I_{2\{111\}}
\end{align}
We substitute this in the reconstruction formula \eqref{reco} to get $\langle O_3O_3J\rangle$. Acting with $D^{++}_{12}$ on $\langle O_3O_3 J\rangle $ gives :
\begin{align}
 \langle JJJ\rangle &= D^{++}_{12} \langle O_3O_3 J\rangle +\text{cyclic perms.}\nonumber\\[5pt]
 &=  D^{++}_{12}\langle O_3O_3 j\rangle +\text{cyclic perms.}\label{WSA}
\end{align}
since $D^{++}_{12}$ kills the longitudinal part. We now compute \eqref{WSA} to get the following explicit form of the form factor :
\begin{align}
&C^{abc}(k_1, k_2, k_3) =\notag\\[5pt] &-\frac{4\,f^{abc}}{J^2}\big[-((k_3\cdot k_1)^2-k_3\cdot k_1 k^2_2)I_{3\{101\}}+((k_2\cdot k_3)^2+(k_3\cdot k_1)(k^2_1+k^2_2))I_{3\{110\}}\nonumber\\[5pt]
&\hspace{2cm}+(-k_2\cdot k_3\,k^2_1+k^2_2\,k^2_3)I_{3\{011\}}\big]
\end{align}
where $J^2 = \left(k_{1}+k_{2}-k_{3}\right)\left(k_{1}-k_{2}+k_{3}\right)\left(-k_{1}+k_{2}+k_{3}\right)\left(k_{1}+k_{2}+k_{3}\right)$.


\section{Discussion}\label{FDir}
In this paper we computed momentum space parity-odd 3-point functions of scalar operators and conserved currents in a CFT. We used conformal Ward identities to fix the parity-odd part of the $\langle JJO\rangle$ correlator. While using this method to fix more complicated 3-point functions of spinning operators such as $\langle JJJ\rangle_{\text{odd}}$ or $\langle TTT\rangle_{\text{odd}}$, we faced the difficulty of identifying independent tensor structures after the application of the generator of special conformal transformations on the ansatz for the correlator. The difficulty arises from Schouten identities that relate the various tensor structures. We leave this problem to a future work.

 
We defined parity-odd spin-raising and weight-shifting operators, and used them to compute $\langle JJJ\rangle_{\text{odd}}$ and $\langle TTO_{\Delta}\rangle_{\text{odd}}$ for $\Delta=1,2$. However, we found it quite complicated to generalise this analysis to obtain correlators such as $\langle JJT\rangle_{\text{odd}}$ and $\langle TTT\rangle_{\text{odd}}$. The difficulty arises from the large number of possible paths to reach a specific correlator starting from a seed correlator. Another difficulty was the difference in the singularity structure of the seed correlator and the correlator of interest to us. We leave further investigation for the future.

Some of the directions that we would like to pursue in the near future are the following. 
It will be interesting to extend the analysis of \cite{Farrow:2018yni,Lipstein:2019mpu} and study the double copy structure of parity-odd correlation functions \cite{wip}. 
It will be interesting to generalise the construction of momentum space conformal blocks in \cite{Gillioz:2020wgw} to the case where parity-odd contributions are important. In momentum space, conformal blocks are constructed quite simply by taking products of 3-point functions of primary operators (contrasted against position space, where an infinite sum over conformal descendants is required) \cite{Gillioz:2019lgs, gillioz2019convergent}. 
Recently, in the study of cosmological correlators \cite{Arkani-Hamed:2018kmz, Baumann:2019oyu, Baumann:2020dch}, the form of tree-level four-point functions in momentum space was constrained. Tree level spinning correlators  such as $\langle JJOO\rangle$ can get parity-odd contributions as they are built out of products of three-point functions. It would be interesting to find the explicit form of these and study their physical implications. Another interesting direction to pursue is to understand the parity-odd structure of 3-point correlators with operators of arbitrary spin. To do this, 
in addition to the techniques used in this paper, one could use the constraints imposed on the correlators by higher spin equations \cite{Jain:2020rmw,Jain:2020puw}. Solving higher spin equations requires us to include possible contact terms in the correlator. It would be interesting to classify both parity-even and parity-odd contact terms for a given spinning correlator. One can also study momentum space correlation functions of spinning operators in supersymmetric theories \cite{Aharony:2019mbc,Inbasekar:2019wdw}.

\section*{Acknowledgments}
We thank Nilay Kundu and Vinay Malvimat for discussions and collaboration at an early stage. The work of SJ and RRJ is supported by the Ramanujan Fellowship. AM would like to acknowledge the support of CSIR-UGC
(JRF) fellowship (09/936(0212)/2019-EMR-I). The work of AS is supported by the KVPY scholarship.
We acknowledge our debt to the people of India for their steady support of research in basic sciences.

\appendix
\section{Embedding space formalism}
\label{embeddingspaceapp}
In this section we briefly review some aspects of the embedding space formalism following \cite{Costa:2011mg}. Conformal invariance is most manifest in the embedding space formalism as the $d$-dimensional Euclidean conformal algebra is the $(d+2)$-dimensional Poincare algebra. The $d$-dimensional CFT correlator in position space can be written in terms of $(d+2)$-dimensional embedding space. The embeddding space coordinates $X^M$  are defined in terms of the position space coordinates $x^{i}$ as follows :
\begin{align}
X^A&=(X^{+}, X^{-}, X^{i})=(1, x^{2}, x^{i})\nonumber\\
Z^B&=(Z^{+}, Z^{-}, Z^{i})=(0, 2z\cdot x, z^{i})\nonumber\\
X_A&=(X_{+}, X_{-}, X_{i})=(x^2/2,1/2, -x^{i})\nonumber\\
Z_B&=(Z_{+}, Z_{-}, Z_{i})=(z\cdot x, 0, -z^{i})
\end{align}
where the $Z^M$ and $z^i$ are the null polarization vectors in the embedding space and position space, respectively. The derivative on the embedding space is defined as follows
\begin{align}
\frac{\partial}{\partial X^M} = -D\delta^{+}_M + P_i\delta_{M}^i 
\end{align}
where
\begin{align}
D= \Delta+x^i\frac{\partial}{\partial x^i},\quad P_i = \frac{\partial}{\partial x^i}
\end{align}
The $(d+2)$-dimensional space metric is as follows
\begin{align}
\eta_{MN} = 
\begin{pmatrix}
0 && \frac{1}{2} && 0 && 0 && 0\\
\frac{1}{2} && 0 && 0 && 0 && 0\\
0 && 0 && -1 && 0 && 0\\
0 && 0 && 0 && -1 && 0\\
0 && 0 && 0 && 0 && -1\\
\end{pmatrix}
\end{align}
The embedding space coordinates are defined such that $Z\cdot X = X^2 = Z^2 = 0$ because the CFT is defined on the null light cone of the AdS space.

The spin-raising and weight-shifting operators that we study in this paper are constructed to have two properties in the embedding space, namely transversality and interiority. Transversality requires that under the transformation $Z_i \to Z_i +\beta X_i$, the operators remain invariant, while under interiority, the operator must map null light cone to itself. The operators thus constructed are manifestly invariant under conformal transformations.

In the main text we have often used notations such as $\epsilon(Z_1,Z_2,X_1,X_2,X_3)$ :
\begin{align}
\epsilon(Z_1,Z_2,X_1,X_2,X_3)=\epsilon^{ABCDE}Z_{1A}Z_{2B}X_{1C}X_{2D}X_{3E}\,.
\end{align}
%
	
\section{Parity-odd two-point functions}\label{2pt}
As is well known, scale invariance completely fixes CFT two-point functions. parity-odd structures can exist for two-point functions of spinning operators. 
	\subsection{Four and Higher dimensions}
	In four or higher dimensions, it is not possible to have any parity-odd two-point function of  either spin-one or any other spinning symmetric spinning correlator. This is because a parity-odd correlator must necessarily involve the $\epsilon$ tensor and it is simple to show that it is impossible to have any parity-odd 2-point function of a spin-1 or any symmetric tensor operator.
	\subsection{Three-dimensions}
	         In three-dimensions parity-odd two-point functions exist. These come from purely contact terms \footnote{In this case, the corresponding position space correlator with separated points vanishes.}.
	We will look at the parity-odd 2-point functions of spin-one and spin two conserved currents.
	\subsection*{$\langle J^{\mu}J^{\nu}\rangle_{\text{odd}}$}
The general ansatz for the correlator is given by
\begin{equation}
\langle J^{\mu}(k)J^{\nu}(-k)\rangle_{\text{odd}}=A(k)\epsilon^{\mu\nu k}
\end{equation}
The ansatz guarantees that the correlator is transverse to the momentum. Imposing scale invariance gives the following differential equation for the form factor $A(k)$ :
\begin{equation}
k\frac{\partial}{\partial k}A(k)=0
\end{equation}
This implies that the form factor is just a constant in this case and we have :
\begin{equation}
\langle J^{\mu}(k)J^{\nu}(-k)\rangle_{\text{odd}}=c_J \epsilon^{\mu\nu k}
\end{equation}
We will now consider the parity-odd 2-point function of the stress-tensor.
\subsection*{$\langle T^{\mu\nu}T^{\rho\sigma}\rangle_{\text{odd}}$}
We consider the following ansatz for this correlator :
\begin{equation}
\langle T^{\mu\nu}(k)T^{\rho\sigma}(-k)\rangle_{\text{odd}}=B(k)\Delta^{\mu\nu\rho\sigma}(k)
\end{equation}
where $\Delta^{\mu\nu\rho\sigma}(k)$ is a parity-odd, transverse-traceless projector given by :
\begin{equation}
\Delta^{\mu\nu\rho\sigma}(k)=\epsilon^{\mu\rho k}\pi^{\nu\sigma}(k)+\epsilon^{\mu\sigma k}\pi^{\nu\rho}(k)+\epsilon^{\nu\sigma k}\pi^{\mu\rho}(k)+\epsilon^{\nu\rho k}\pi^{\mu\sigma}(k)
\end{equation}
where $\pi^{\mu}_{\nu}(k)$ is the same projector used in previous sections. The ansatz guarantees that the correlator is transverse and traceless.

\noindent The dilatation Ward identity gives the following equation for the form factor $B(k)$ :
\begin{equation}
\left(k\frac{\partial}{\partial k}-2\right)B(k)=0
\end{equation}
This can be easily solved to get
\begin{equation}
B(k)=c_T k^2
\end{equation}
Therefore, the correlator is given by
\begin{equation}
\langle T^{\mu\nu}(k)T^{\rho\sigma}(-k)\rangle_{\text{odd}}=c_T \Delta^{\mu\nu\rho\sigma}(k)k^2
\end{equation}


\section{Schouten Identities}\label {schouten}
Here, we list the Schouten identities used in our calculations in the main text. The most general form of a Schouten identity in $d$-dimensions is 
\begin{equation}
\epsilon^{\left[\mu_1\mu_2 \ldots \mu_d\right.}\delta^{\left.\nu\right]\rho}=0
\end{equation}
In three-dimensions, this translates to
\begin{equation}\label{schouten3D}
\epsilon^{\mu_1\mu_2\mu_3}\delta^{\nu\rho}-\epsilon^{\mu_2\mu_3\nu}\delta^{\mu_1\rho}+\epsilon^{\mu_3\nu\mu_1}\delta^{\mu_2\rho}-\epsilon^{\nu\mu_1\mu_2}\delta^{\mu_3\rho}=0
\end{equation}
Dotting the indices in \eqref{schouten3D} with momenta lets us relate different epsilon structures that occur in the correlation functions calculated earlier. Dotting with $k_{1\mu_3}$, $k_{1\rho}$ and $k_{2\nu}$ gives
\begin{align}\label{jjoschouten1}
\epsilon^{\mu_1\mu_2 k_1}(k_1\cdot k_2)+\epsilon^{\mu_1 k_1 k_2}k_1^{\mu_2}&=\epsilon^{\mu_1\mu_2 k_2}k_1^2+\epsilon^{\mu_2 k_1 k_2}k_1^{\mu_1}
\end{align}
Similarly, dotting with $k_{1\mu_3}$, $k_{2\rho}$ and $k_{2\nu}$
\begin{align}\label{jjoschouten2}
\epsilon^{\mu_1\mu_2 k_2}(k_1\cdot k_2)+\epsilon^{\mu_2 k_1 k_2}k_2^{\mu_1}=\epsilon^{\mu_1\mu_2 k_1}k_2^2+\epsilon^{\mu_1 k_1 k_2}k_2^{\mu_2}
\end{align}
\eqref{jjoschouten1} and \eqref{jjoschouten2} were useful in rewriting the $\langle JJO \rangle$ ansatz. One can also derive these two identities by considering the contraction of three Levi-Civita tensors.
\begin{align}
\epsilon^{\mu_1\alpha k_1}\epsilon^{\beta k_2\mu_2}\epsilon_{\beta\rho\alpha}&=\epsilon^{\mu_1\mu_2 k_1}k_{2\rho}+\epsilon^{\mu_1 k_1 k_2}\delta^{\mu_2}_\rho\cr
\epsilon^{\beta k_2\mu_2}\epsilon^{\alpha k_1 \mu_1}\epsilon_{\alpha\beta\rho}&=\epsilon^{\mu_1\mu_2 k_2}k_{1\rho}+\epsilon^{\mu_2 k_1 k_2}\delta^{\mu_1}_\rho
\end{align}
Equating the RHS of the two equations after dotting them with $k_1^\rho$ and $k_2^\rho$ respectively, we get back \eqref{jjoschouten1} and \eqref{jjoschouten2}. Similarly, while checking the transverse identity for $\langle JJJ \rangle$, we used the following Schouten identities
\begin{align}\label{jjjschouten1}
\epsilon^{k_1 k_2 \mu_3}k_2^{\mu_2}&=\epsilon^{k_2 \mu_3\mu_2}(k_1\cdot k_2)+\epsilon^{\mu_2 \mu_3 k_1}k_2^2+\epsilon^{k_1 k_2 \mu_2}k_2^{\mu_3}\\
\label{jjjschouten2}
\epsilon^{k_1 k_2 \mu_3}k_3^{\mu_2}&=\epsilon^{k_2 \mu_3\mu_2}(k_1\cdot k_3)+\epsilon^{\mu_2\mu_3 k_1}(k_2\cdot k_3)+\epsilon^{\mu_3 k_1 k_2}k_3^{\mu_3}
\end{align}
In four-dimensions, we use the following identities to rewrite the ansatz for $\langle JJJ \rangle$. 
\begin{align}\label{schouten4D}
&(k_1 \cdot k_2) \epsilon_{\mu_{1} \mu_2 \mu_3 k_1}-k_{1 \mu_1}\epsilon_{k_2 \mu_2 \mu_3 k_1}-k_{1 \mu_2} \epsilon_{\mu_1 k_2 \mu_3 k_1}-k_{1 \mu_3} \epsilon_{\mu_1\mu_2 k_2 k_1}-k^2_1\epsilon_{\mu_1\mu_2 \mu_3 k_2 }=0\notag\\
&(k_1 \cdot k_2) \epsilon_{\mu_{1} \mu_2 \mu_3 k_2}-k_{2 \mu_1}\epsilon_{k_2 \mu_2 \mu_3 k_2}-k_{2 \mu_2} \epsilon_{\mu_1 k_1 \mu_3 k_2}-k_{2 \mu_3} \epsilon_{\mu_1\mu_2 k_1 k_2}-k^2_2\epsilon_{\mu_1\mu_2 \mu_3 k_1 }=0
\end{align}


\section{Computation details for $\langle TTO\rangle$}\label{CD}
In this section we give some details of the computation of the $\langle TTO\rangle$ correlator in the free fermion theory \eqref{FFaction}. We use the form of the stress tensor and the scalar operator as given in \eqref{op} and \eqref{op11}.
The correlator of interest in terms of spinor fields is as follows :
\begin{align}
&\langle T_{\mu\nu}(k_1)T_{\alpha\beta}(k_2)O(k_3)\rangle \notag\\&= \int_{123} \langle \bar{\psi}(l_1)\gamma_{\mu}(2l_1-k_1)_{\nu}\psi(k_1-l_1)\bar{\psi}(l_2)\gamma_{\alpha}(2l_2-k_2)_{\beta}\psi(k_2-l_2)\bar{\psi}(l_3)\psi(k_3-l_3)\rangle
\end{align}
The Wick contraction $(1\bar 3) (3 \bar 2) (2 \bar 1) $ gives \footnote{We are not giving the details of the complex conjugate here as it gives the same result.} :
\begin{align}
&\int_{123} tr(\gamma_{\mu}\gamma_{\rho}\gamma_{\sigma}\gamma_{\alpha}\gamma_{\tau})(2l_1-k_1)_{\nu}(2l_2-k_2)_{\beta}\frac{l^{\rho}_3l^{\sigma}_2l^{\tau}_1}{l^2_1l^2_2l^2_3}\delta(k_1-l_1+l_3)\delta(k_3-l_3+l_2)\delta(k_2-l_2+l_1)\label{Wick}
\end{align}
where the fermion propagator given by :
\begin{align}
\langle\bar{\psi}_{\alpha}(k_1)\psi_{\beta}(k_2)\rangle = \delta^{(3)}(k_1+k_2)\frac{\slashed{k}_{1,\alpha\beta}}{k^2_1}
\end{align}
was used. Computing the integrals over $l_1, l_2$ and using momentum conservation, we may write the integral \eqref{Wick} over a single variable as :
\begin{align}
&\frac{1}{4}\left[\int_3 tr(\gamma_{\tau}\gamma_{\sigma}\gamma_{\alpha}\gamma_{\rho}\gamma_{\mu})(2l_3-k_1)_{\nu}(2l_3+k_2)_{\beta}\frac{l_3^{\rho}(l_3-k_1)^{\tau}(l_3+k_2)^{\sigma}}{l_3^2(l_3+k_2)^2(l_3-k_1)^2}\right]\notag\\[5pt]&+\text{symmetrize in $(\mu, \nu)$ and $(\alpha, \beta)$}\label{int}
\end{align}
By projecting \eqref{int} with spin-2 projectors, one can now determine the transverse part of the correlator and hence, identify the form factors in \eqref{ttoform} explicitly.
\subsection{Transverse and trace Ward identities}\label{WI}
Here we provide an explicit verification of the transverse and trace Ward identities satisfied by $\langle TTO\rangle$. Contracting \eqref{int} with $k_{1\mu}$ we obtain :
\begin{align}
&\int_3 tr(\frac{1}{\slashed{l}_3-\slashed{k}_1}\frac{1}{\slashed{l}_3+\slashed{k}_2}\gamma_{\alpha}\frac{1}{\slashed{l}_3}\slashed{k}_1)(2l_3-k_1)_{\nu}(2l_3+k_2)_{\beta}\notag\\[5pt]
&+\int_3 tr(\frac{1}{\slashed{l}_3-\slashed{k}_1}\frac{1}{\slashed{l}_3+\slashed{k}_2}\gamma_{\alpha}\frac{1}{\slashed{l}_3}\gamma_{\mu})(2l_3-k_1)\cdot k_1(2l_3+k_2)_{\beta}\notag\\[5pt]&+\text{symmetrize in $(\alpha, \beta)$}
\end{align}
which when simplified using :
\begin{align}
&\frac{1}{\slashed{l}_3}\slashed{k}_1\frac{1}{\slashed{l}_3-\slashed{k}_1}=\frac{1}{\slashed{l}_3-\slashed{k}_1}-\frac{1}{\slashed{l}_3}
\notag\\[5pt]
&(2l_3-k_1)\cdot k_1 =- (l_3-k_1)^2+l^2_3
\end{align}
After a little algebra we get
\begin{align}
\frac{1}{128}k_3\big [k_{1\beta}(\epsilon_{\alpha\nu k_1}-\epsilon_{\alpha\nu k_2})+k_{1\alpha}(\epsilon_{\beta\nu k_1}-\epsilon_{\beta\nu k_2}
)-\delta_{\alpha, \nu}\epsilon_{\beta k_1 k_2}-\delta_{\beta, \nu}\epsilon_{\alpha k_1 k_2}\big ] \label{Twi}
\end{align}
which precisely reproduces the transverse Ward identity given in \eqref{fdWI}. We now contract the $(\mu, \nu)$ indices to check the trace Ward identity :
\begin{align}
&\int_3 tr(\frac{1}{\slashed{l}_3-\slashed{k}_1}\frac{1}{\slashed{l}_3+\slashed{k}_2}\gamma_{\alpha}\frac{1}{\slashed{l}_3}(2\slashed{l}_3-\slashed{k}_1))(2l_3+k_2)_{\beta}\notag\\& =\int_3 tr(\frac{1}{\slashed{l}_3-\slashed{k}_1}\frac{1}{\slashed{l}_3+\slashed{k}_2}\gamma_{\alpha})(2l_3+k_2)_{\beta}+\int_3 tr(\frac{1}{\slashed{l}_3}\frac{1}{\slashed{l}_3+\slashed{k}_2}\gamma_{\alpha})(2l_3+k_2)_{\beta}
\end{align}
In the first term, one may transform the integration variable to $l_3\to l_3-k_2$ :
\begin{align}
&\int_3 tr(\frac{1}{\slashed{l}_3+\slashed{k}_3}\frac{1}{\slashed{l}_3}\gamma_{\alpha})(2l_3+k_2)_{\beta}+\int_3 tr(\frac{1}{\slashed{l}_3}\frac{1}{\slashed{l}_3+\slashed{k}_2}\gamma_{\alpha})(2l_3+k_2)_{\beta}\notag\\& = -2i[\int_3\frac{\epsilon^{l_3 k_3\alpha}}{l^2_3(l_3+k_3)^2}(2l_3-k_2)_{\beta}-\int_3\frac{\epsilon^{l_3 k_2\alpha}}{l^2_3(l_3+k_2)^2}(2l_3+k_2)_{\beta}] \label{trint}
\end{align}
Making use of the following identities
\begin{align}
&\int_l \frac{l_\mu}{l^2(l+k)^2} = A k_\mu
\notag\\
&\int_l \frac{l_\mu l_{\nu}}{l^2(l+k)^2} = B\eta_{\mu\nu}+C k_\mu k_\nu
\end{align}
where $A, B, C$ are scalars, one can see that \eqref{trint} vanishes and hence show that the correlator satisfies \eqref{TrWIE}.
\subsection{Details of longitudinal part}
\label{TTO2long}
In this section we compute the function derivative $\frac{\delta T_{\mu\nu}(x)}{\delta g_{\alpha\beta}(y)}$ in the free fermion theory, relevant to Section \ref{TTO2FFsection}.
Since the functional dependence on the metric is only via the spin connection we have :
\begin{align}
\frac{\delta T_{\mu\nu}(x)}{\delta g_{\alpha\beta}(y)}\bigg|_{g_{\alpha\beta} \to \eta_{\alpha\beta}} &=\frac{1}{32}\bigg(\frac{ \delta\omega^{~ab}_{\nu}(x)}{\delta g_{\alpha\beta}(y)}|_{g_{\alpha\beta} \to \eta_{\alpha\beta}}\bar{\psi}\{\gamma_{\mu},\gamma_{ab}\}\psi+\frac{ \delta\omega^{~ab}_{\mu}(x)}{\delta g_{\alpha\beta}(y)}|_{g_{\alpha\beta} \to \eta_{\alpha\beta}}\bar{\psi}\{\gamma_{\nu},\gamma_{ab}\}\psi\bigg)
\notag\\[5pt]&=\frac{i}{16}\left(\frac{\delta\omega^{~ab}_{\nu}(x)}{\delta g_{\alpha\beta}(y)}\epsilon_{~ab\mu}+\frac{\delta\omega^{~ab}_{\mu}(x)}{\delta g_{\alpha\beta}(y)}\epsilon_{ab\nu}\right)O(x) \label{fd}
\end{align}
where
\begin{align}
\frac{ \delta\omega^{~ab}_{\nu}(x)}{\delta g_{\alpha\beta}(y)}|_{g_{\alpha\beta} \to \eta_{\alpha\beta}}= \frac{\delta^{\sigma b}\delta^{a\tau}}{2}\left[\partial_{\sigma}(\delta^{\alpha}_{(\tau}\delta^{\beta}_{\nu)}\delta^{(3)}(x-y))+\partial_{\nu}(\delta^{\alpha}_{(\tau}\delta^{\beta}_{\sigma)}\delta^{(3)}(x-y))-\partial_{\tau}(\delta^{\alpha}_{(\sigma}\delta^{\beta}_{\nu)}\delta^{(3)}(x-y))\right]
\end{align}
The second line in \eqref{fd} was obtained by recognising that in three-dimensions, $\gamma_{ab} = [\sigma_a, \sigma_b] = 2i\epsilon_{abc}\sigma^c$. In the limit,  $g_{\alpha\beta} \to \eta_{\alpha\beta}$, the vierbiens go to $e^{a}_{\alpha}\to \delta^a_{\alpha}$ was also used.  Simplifying the above expression to obtain,
\begin{align}
\frac{ \delta\omega^{~ab}_{\nu}(x)}{\delta g_{\alpha\beta}(y)}\epsilon_{ab\mu} =-[\epsilon_{\sigma\alpha\mu}\delta_{\beta\nu}+\epsilon_{\sigma\beta\mu}\delta_{\alpha\nu}]\partial^{\sigma}\delta^{(3)}(x-y) 
\end{align}
From here we get \eqref{TTO2int}.

\section{Parity-even spin-raising and weight-shifting operators}\label{ParityEvenOperators}
In this section we list out all the parity-even weight-shifting operators used in the main text of the paper \cite{Baumann:2019oyu,Baumann:2020dch}.

The operator that decreases the scaling dimension of operators at points 1 and 2 is :
\begin{align}\label{W12minus}
W_{12}^{--}&=\frac{1}{2}\vec{K}^-_{12}\cdot\vec{K}^-_{12}
\end{align}
where 
\begin{align}
K_{12}^{-\mu}&=\partial_{k_{1\mu}}-\partial_{k_{2\mu}}
\end{align}
We also use 
\begin{align}
K_{12}^{+\mu}&=\partial_{k_{1\mu}}+\partial_{k_{2\mu}}
\end{align}
We can also define an operator that increases the scaling dimension at 2-points. Although this has a very complicated expression, it simplifies when acting on scalar operators and is given by :
\begin{align}\label{W12plus}
\begin{split}
W_{12}^{++}=&(k_1 k_2)^2\,W_{12}^{--}-(d-2\,\Delta_1)(d-2\,\Delta_2)k_1\cdot k_2\\[5pt]
&+\left(k_2^2(d-2\,\Delta_1)(d-1-\Delta_1+k_1\cdot K_{12})+(1 \leftrightarrow 2)\right)
\end{split}
\end{align}
$D_{11}$ raises the spin of the operator at point 1 and simultaneously lowers its weight. This was used in the construction of both $\langle TTO\rangle$ and $\langle JJJ\rangle$ :
\begin{align}\label{D11}
D_{11}=&(\Delta_2-3+\vec{k}_2\cdot\vec{K}_{12})\vec{z}_1\cdot\vec{K}_{12}-(\vec{k}_2\cdot\vec{z}_1)W_{12}^{--}-(\vec{z}_2\cdot\vec{K}_{12})\,(\vec{z}_1\cdot\partial_{\vec{z}_2})+(\vec{z}_1\cdot\vec{z}_2)\partial_{\vec{z}_2}\cdot\vec{K}_{12}
\end{align}
 We can similarly define $D_{22}$ and $D_{33}$ by doing cyclic permutations of the momenta and polarization vectors in \eqref{D11}. For example,
\begin{equation}\label{D22}
D_{22}\left((k_1, z_1), (k_2, z_2), (k_3, z_3)\right)=D_{11}\left((k_3, z_3), (k_1, z_1), (k_2, z_2)\right)
\end{equation}
$S_{12}^{++}$ raises the spin at points 1 and 2 :
\begin{align}\label{S12plus}
\begin{split}
S_{12}^{++}=&(s_1+\Delta_1-1)(s_2+\Delta_2-1)z_1\cdot z_2-(z_1\cdot k_1)(z_2\cdot k_2)W_{12}^{--}\\[6 pt]
&+\left[(s_1+\Delta_1-1)(k_2\cdot z_2)(z_1\cdot K_{12})+\left(1 \leftrightarrow 2\right)\right]
\end{split}
\end{align}
$S_{23}^{++}$ and $S_{13}^{++}$ are once again defined by cyclic permutations of \eqref{S12plus}. 

The operator $H_{12}$ which raises the spin at points 1 and 2 and also lowers the weight at both the points is given by :
\begin{equation}\label{H12}
H_{12}=2\,(z_1\cdot K_{12})(z_2\cdot K_{12})-2\,(z_1\cdot z_2)W_{12}^{--}
\end{equation}
The operator that raises the spin at point 1 and simulataneously lowers the weight at point 2 is given by :
\begin{equation}\label{D12}
D_{12}=(\Delta_1+s_1-1)z_1\cdot K_{12}-(z_1\cdot k_1)W_{12}^{--}
\end{equation}
A $(1 \leftrightarrow 2)$ exchange in this operator gives $D_{21}$. Both of these were used in the construction of $\langle TTO\rangle$. 


\section{Embedding space parity-odd correlation functions in four-dimensions}\label{TTT}
In this appendix, we show  that $\langle J  J  T \rangle_\text{{odd}}$ and $\langle T  T  T \rangle_\text{{odd}}$ are zero in four-dimensions.
\subsection{$\langle JJT \rangle_\text{{odd}}$} 
We first write the  $\langle J  J  T \rangle_\text{{odd}}$ correlator in a basis of (embedding space) conformally invariant structures (the notation used is that of \cite{Costa:2011mg})
\begin{align}
&\langle J(Z_1, X_1)J(Z_2, X_2)T(Z_3, X_3)\rangle_\text{{odd}} = c_1\,\epsilon(Z_1 Z_2 Z_3 X_1 X_2 X_3)\frac{V_3}{X^{3/2}_{12}X^{5/2}_{23}X^{5/2}_{31}}
\end{align}
Under simultaneous exchange of $(X_1, X_2)$ and $(Z_1, Z_2)$ we have
\begin{align}
V_3 \to -V_3
\end{align}
Hence, symmetry consideration demands that $c_1 = 0$.
This implies that the parity-odd $\langle J  J  T \rangle$  correlator vanishes in four-dimensions. In momentum space, it can be a contact term.
\subsection{$\langle TTT\rangle_\text{{odd}}$}
We first write the correlator in a basis of conformally invariant structures
\begin{align}
&\langle T(Z_1, X_1)T(Z_2, X_2)T(Z_3, X_3)\rangle_\text{{odd}} \notag\\[5pt]&=\epsilon(Z_1 Z_2 Z_3 X_1 X_2 X_3) \sum_{n_{12}, n_{13}, n_{23}}A_{n_{12}, n_{13}, n_{23}}\frac{V^{1-n_{12}-n_{13}}_1V^{1-n_{12}-n_{23}}_2V^{1-n_{13}-n_{23}}_3 H^{n_{12}}_{12} H^{n_{13}}_{13}H^{n_{23}}_{23}}{X^{7/2}_{12}X^{7/2}_{13}X^{7/2}_{23}}\notag\\[5pt]&=\epsilon(Z_1 Z_2 Z_3 X_1 X_2 X_3)\notag\\[5pt]&[A_{000}\frac{V_1V_2V_3}{X^{7/2}_{12}X^{7/2}_{13}X^{7/2}_{23}}+A_{001}\frac{V_2  H_{13}}{X^{7/2}_{12}X^{7/2}_{13}X^{7/2}_{23}}+A_{010}\frac{V_1  H_{23}}{X^{7/2}_{12}X^{7/2}_{13}X^{7/2}_{23}}+A_{100}\frac{V_3  H_{12}}{X^{7/2}_{12}X^{7/2}_{13}X^{7/2}_{23}}]
\end{align}
Under simultaneous $Z_1 \leftrightarrow Z_2$, $X_1 \leftrightarrow X_2$ exchange we have
\begin{align}
V_1 \to -V_2, \quad V_3 \to -V_3, \quad H_{12} \to H_{12}, \quad H_{13} \to  H_{23}
\end{align}
\begin{align}
&\langle T(Z_1, X_1)T(Z_2, X_2)T(Z_3, X_3)\rangle_\text{{odd}} \notag\\[5pt]&=-\epsilon(Z_1 Z_2 Z_3 X_1 X_2 X_3)\notag\\[5pt]&[A_{000}\frac{V_1V_2V_3}{X^{7/2}_{12}X^{7/2}_{13}X^{7/2}_{23}}+A_{001}\frac{V_1  H_{23}}{X^{7/2}_{12}X^{7/2}_{13}X^{7/2}_{23}}+A_{010}\frac{V_2  H_{13}}{X^{7/2}_{12}X^{7/2}_{13}X^{7/2}_{23}}+A_{100}\frac{V_3  H_{12}}{X^{7/2}_{12}X^{7/2}_{13}X^{7/2}_{23}}]
\end{align}
Therefore, we must have
\begin{align}
A_{000} = 0, \quad A_{001} = -A_{010}, \quad A_{100} = 0
\end{align}
Hence, 
\begin{align}
&\langle T(Z_1, X_1)T(Z_2, X_2)T(Z_3, X_3)\rangle_\text{{odd}} =A_{001}\epsilon(Z_1 Z_2 Z_3 X_1 X_2 X_3)\left[\frac{V_1  H_{23}}{X^{7/2}_{12}X^{7/2}_{13}X^{7/2}_{23}}-\frac{V_2  H_{13}}{X^{7/2}_{12}X^{7/2}_{13}X^{7/2}_{23}}\right]
\end{align}
Now, under simultaneous exchange of $(Z_2, Z_3)$ and $(X_2, X_3)$
\begin{align}
V_{1} \to -V_1 \quad V_{2} \to -V_3 \quad H_{23} \to H_{23} \quad H_{13} \to H_{12}
\end{align}
Therefore,
\begin{align}
&\langle T(Z_1, X_1)T(Z_2, X_2)T(Z_3, X_3)\rangle_\text{{odd}} =A_{001}\epsilon(Z_1 Z_2 Z_3 X_1 X_2 X_3)\left[-\frac{V_1  H_{23}}{X^{7/2}_{12}X^{7/2}_{13}X^{7/2}_{23}}+\frac{V_3  H_{12}}{X^{7/2}_{12}X^{7/2}_{13}X^{7/2}_{23}}\right]
\end{align}
Hence, we must have $A_{001} = 0$. Therefore, symmetry considerations force $\langle TTT\rangle_\text{{odd}}$ to be zero.

\bibliographystyle{JHEP}

\begin{thebibliography}{101}

\bibitem{Coriano:2013jba}
C.~Coriano, L.~Delle Rose, E.~Mottola and M.~Serino,
``Solving the Conformal Constraints for Scalar Operators in Momentum Space and the Evaluation of Feynman's Master Integrals,''
JHEP \textbf{07} (2013), 011
[\href{https://arxiv.org/abs/1304.6944}{{\tt arXiv:1304.6944   [hep-th]}}].

\bibitem{Bzowski:2013sza}
A.~Bzowski, P.~McFadden and K.~Skenderis,
``Implications of conformal invariance in momentum space,''
JHEP \textbf{03} (2014), 111
[\href{https://arxiv.org/abs/1304.7760}{{\tt arXiv:1304.7760 [hep-th]}}].

\bibitem{Bonora:2015nqa}
L.~Bonora, A.~D.~Pereira and B.~Lima de Souza,
``Regularization of energy-momentum tensor correlators and parity-odd terms,''
JHEP \textbf{06} (2015), 024
[\href{https://arxiv.org/abs/1503.03326}{{\tt arXiv:1503.03326 [hep-th]}}].







\bibitem{Bzowski:2015pba}
A.~Bzowski, P.~McFadden and K.~Skenderis,
``Scalar 3-point functions in CFT: renormalisation, beta functions and anomalies,''
JHEP \textbf{03} (2016), 066
[\href{https://arxiv.org/abs/1510.08442}{{\tt arXiv:1510.08442 [hep-th]}}].




\bibitem{Bzowski:2015yxv}
A.~Bzowski, P.~McFadden and K.~Skenderis,
``Evaluation of conformal integrals,''
JHEP \textbf{02} (2016), 068
[\href{https://arxiv.org/abs/1511.02357}{{\tt arXiv:1511.02357 [hep-th]}}].

\bibitem{Bonora:2015odi}
L.~Bonora and B.~Lima de Souza,
``Pure contact term correlators in CFT,''
Bled Workshops Phys. \textbf{16} (2015) no.2, 22-34
[\href{https://arxiv.org/abs/1511.06635}{{\tt arXiv:1511.06635 [hep-th]}}].



\bibitem{Bonora:2016ida}
L.~Bonora, M.~Cvitan, P.~Dominis Prester, B.~Lima de Souza and I.~Smolić,
``Massive fermion model in 3d and higher spin currents,''
JHEP \textbf{05} (2016), 072
[\href{https://arxiv.org/abs/1602.07178}{{\tt arXiv:1602.07178 [hep-th]}}].



\bibitem{sissathesis}
B.~Lima de Souza,
``CFT’s, contact terms and anomalies"
PhD Thesis


\bibitem{Bzowski:2017poo}
A.~Bzowski, P.~McFadden and K.~Skenderis,
``Renormalised 3-point functions of stress tensors and conserved currents in CFT,''
JHEP \textbf{11} (2018), 153
[\href{https://arxiv.org/abs/1711.09105}{{\tt arXiv:1711.09105 [hep-th]}}].



\bibitem{Coriano:2018bbe}
C.~Corianò and M.~M.~Maglio,
``Exact Correlators from Conformal Ward Identities in Momentum Space and the Perturbative $TJJ$ Vertex,''
Nucl. Phys. B \textbf{938} (2019), 440-522
[\href{https://arxiv.org/abs/1802.07675}{{\tt arXiv:1802.07675 [hep-th]}}].


\bibitem{Isono:2018rrb}
H.~Isono, T.~Noumi and G.~Shiu,
``Momentum space approach to crossing symmetric CFT correlators,''
JHEP \textbf{07} (2018), 136
[\href{https://arxiv.org/abs/1805.11107}{{\tt arXiv:1805.11107 [hep-th]}}].

\bibitem{Bzowski:2018fql}
A.~Bzowski, P.~McFadden and K.~Skenderis,
``Renormalised CFT 3-point functions of scalars, currents and stress tensors,''
JHEP \textbf{11} (2018), 159
[\href{https://arxiv.org/abs/1805.12100}{{\tt arXiv:1805.12100 [hep-th]}}].

\bibitem{Gillioz:2018mto}
M.~Gillioz,
``Momentum-space conformal blocks on the light cone,''
JHEP \textbf{10} (2018), 125
[\href{https://arxiv.org/abs/1807.07003}{{\tt arXiv:1807.07003 [hep-th]}}].

\bibitem{Coriano:2018tgn}
C.~Corianò and M.~M.~Maglio,
``Conformal Ward Identities and the Coupling of QED and QCD to Gravity,''
EPJ Web Conf. \textbf{192} (2018), 00047
[\href{https://arxiv.org/abs/1809.05940}{{\tt arXiv:1809.05940 [hep-th]}}].

\bibitem{Albayrak:2018tam}
S.~Albayrak and S.~Kharel,
``Towards the higher point holographic momentum space amplitudes,''
JHEP \textbf{02} (2019), 040
[\href{https://arxiv.org/abs/1810.12459}{{\tt arXiv:1810.12459 [hep-th]}}].

\bibitem{Farrow:2018yni}
J.~A.~Farrow, A.~E.~Lipstein and P.~McFadden,
``Double copy structure of CFT correlators,''
JHEP \textbf{02}, 130 (2019)
[\href{https://arxiv.org/abs/1812.11129}{{\tt arXiv:1812.11129 [hep-th]}}].



\bibitem{Isono:2019ihz}
H.~Isono, T.~Noumi and T.~Takeuchi,
``Momentum space conformal three-point functions of conserved currents and a general spinning operator,''
JHEP \textbf{05}, 057 (2019)
[\href{https://arxiv.org/abs/1903.01110}{{\tt arXiv:1903.01110[hep-th]}}].

\bibitem{Maglio:2019grh}
C.~Corianò and M.~M.~Maglio,
``On Some Hypergeometric Solutions of the Conformal Ward Identities of Scalar 4-point Functions in Momentum Space,''
JHEP \textbf{09} (2019), 107
[\href{https://arxiv.org/abs/1903.05047}{{\tt arXiv:1903.05047 [hep-th]}}].

\bibitem{Albayrak:2019asr}
S.~Albayrak, C.~Chowdhury and S.~Kharel,
``New relation for Witten diagrams,''
JHEP \textbf{10} (2019), 274
[\href{https://arxiv.org/abs/1904.10043}{{\tt arXiv:1904.10043 [hep-th]}}].

\bibitem{Albayrak:2019yve}
S.~Albayrak and S.~Kharel,
``Towards the higher point holographic momentum space amplitudes. Part II. Gravitons,''
JHEP \textbf{12} (2019), 135
[\href{https://arxiv.org/abs/1908.01835}{{\tt arXiv:1908.01835 [hep-th]}}].

\bibitem{Isono:2019wex}
H.~Isono, T.~Noumi and G.~Shiu,
``Momentum space approach to crossing symmetric CFT correlators. Part II. General spacetime dimension,''
JHEP \textbf{10} (2019), 183
[\href{https://arxiv.org/abs/1908.04572}{{\tt arXiv:1908.04572 [hep-th]}}].

\bibitem{Bautista:2019qxj}
T.~Bautista and H.~Godazgar,
``Lorentzian CFT 3-point functions in momentum space,''
JHEP \textbf{01} (2020), 142
[\href{https://arxiv.org/abs/1908.04733}{{\tt arXiv:1908.04733 [hep-th]}}].




\bibitem{Gillioz:2019lgs}
M.~Gillioz,
``Conformal 3-point functions and the Lorentzian OPE in momentum space,''
Commun. Math. Phys. \textbf{379}, no.1, 227-259 (2020)
[\href{https://arxiv.org/abs/1909.00878}{{\tt arXiv:1909.00878 [hep-th]}}].

\bibitem{Bzowski:2019kwd}
A.~Bzowski, P.~McFadden and K.~Skenderis,
``Conformal $n$-point functions in momentum space,''
Phys. Rev. Lett. \textbf{124} (2020) no.13, 131602
[\href{https://arxiv.org/abs/1910.10162}{{\tt arXiv:1910.10162 [hep-th]}}].





\bibitem{Coriano:2019nkw}
C.~Corianò, M.~M.~Maglio and D.~Theofilopoulos,
``Four-Point Functions in Momentum Space: Conformal Ward Identities in the Scalar/Tensor case,''
[\href{https://arxiv.org/abs/1912.01907}{{\tt arXiv:1912.01907 [hep-th]}}].

\bibitem{gillioz2019convergent}
M.~Gillioz, X.~Lu, M.~A.~Luty and G.~Mikaberidze,
``Convergent Momentum-Space OPE and Bootstrap Equations in Conformal Field Theory,''
JHEP \textbf{03} (2020), 102
[\href{https://arxiv.org/abs/1912.05550}{{\tt arXiv:1912.05550 [hep-th]}}].
%













\bibitem{Lipstein:2019mpu}
A.~E.~Lipstein and P.~McFadden,
``Double copy structure and the flat space limit of conformal correlators in even dimensions,''
Phys. Rev. D \textbf{101}, no.12, 125006 (2020)
[\href{https://arxiv.org/abs/1912.10046}{{\tt arXiv:1912.10046 [hep-th]}}].


\bibitem{Albayrak:2020isk}
S.~Albayrak, C.~Chowdhury and S.~Kharel,
``An étude of momentum space scalar amplitudes in AdS,''
Phys. Rev. D \textbf{101} (2020), 124043
[\href{https://arxiv.org/abs/2001.06777}{{\tt arXiv:2001.06777 [hep-th]}}].

\bibitem{Coriano:2020ccb}
C.~Corianò and M.~M.~Maglio,
``The Generalized Hypergeometric Structure of the Ward Identities of CFT's in Momentum Space in $d > 2$,''
Axioms \textbf{9} (2020), 2, 54
[\href{https://arxiv.org/abs/2001.09622}{{\tt arXiv:2001.09622 [hep-th]}}].

\bibitem{Gillioz:2020mdd}
M.~Gillioz, M.~Meineri and J.~Penedones,
``A Scattering Amplitude in Conformal Field Theory,''
[\href{https://arxiv.org/abs/2003.07361}{{\tt arXiv:2003.07361 [hep-th]}}].

\bibitem{Serino:2020pyu}
M.~Serino,
``The four-point correlation function of the energy-momentum tensor in the free conformal field theory of a scalar field,''
[\href{https://arxiv.org/abs/2004.08668}{{\tt arXiv:2004.08668 [hep-th]}}].

\bibitem{Jain:2020rmw}
S.~Jain, R.~R.~John and V.~Malvimat,
``Momentum space spinning correlators and higher spin equations in three-dimensions ,''
[\href{https://arxiv.org/abs/2005.07212}{{\tt arXiv:2005.07212 [hep-th]}}].

\bibitem{Albayrak:2020bso}
S.~Albayrak and S.~Kharel,
``On spinning loop amplitudes in Anti-de Sitter space,''
[\href{https://arxiv.org/abs/2006.12540}{{\tt arXiv:2006.12540 [hep-th]}}].


\bibitem{Bzowski:2020kfw}
A.~Bzowski, P.~McFadden and K.~Skenderis,
``Conformal correlators as simplex integrals in momentum space,''
[\href{https://arxiv.org/abs/2008.07543}{{\tt arXiv:2008.07543 [hep-th]}}].

\bibitem{Jain:2020puw}
S.~Jain, R.~R.~John and V.~Malvimat,
``Constraining momentum space correlators using slightly broken higher spin symmetry,''
[\href{https://arxiv.org/abs/2008.08610}{{\tt arXiv:2008.08610 [hep-th]}}].


\bibitem{Gillioz:2020wgw}
M.~Gillioz,
``Conformal partial waves in momentum space,''
[\href{https://arxiv.org/abs/2012.09825}{{\tt arXiv:2012.09825 [hep-th]}}].

\bibitem{Albayrak:2020fyp}
S.~Albayrak, S.~Kharel and D.~Meltzer,
``On duality of color and kinematics in (A)dS momentum space,''
[\href{https://arxiv.org/abs/2012.10460}{{\tt arXiv:2012.10460 [hep-th]}}].




\bibitem{Maldacena:2011nz}
J.~M.~Maldacena and G.~L.~Pimentel,
``On graviton non-Gaussianities during inflation,''
JHEP \textbf{09}, 045 (2011)
[\href{https://arxiv.org/abs/1104.2846}{{\tt arXiv:1104.2846 [hep-th]}}].


\bibitem{Mata:2012bx}
I.~Mata, S.~Raju and S.~Trivedi,
``CMB from CFT,''
JHEP \textbf{07} (2013), 015
[\href{https://arxiv.org/abs/1211.5482}{{\tt arXiv:1211.5482 [hep-th]}}].

\bibitem{Ghosh:2014kba}
A.~Ghosh, N.~Kundu, S.~Raju and S.~P.~Trivedi,
``Conformal Invariance and the Four Point Scalar Correlator in Slow-Roll Inflation,''
JHEP \textbf{07} (2014), 011
[\href{https://arxiv.org/abs/1401.1426}{{\tt arXiv:1401.1426 [hep-th]}}].

\bibitem{Kundu:2014gxa}
N.~Kundu, A.~Shukla and S.~P.~Trivedi,
``Constraints from Conformal Symmetry on the 3-point Scalar Correlator in Inflation,''
JHEP \textbf{04} (2015), 061
[\href{https://arxiv.org/abs/1410.2606}{{\tt arXiv:1410.2606 [hep-th]}}].

\bibitem{Arkani-Hamed:2015bza}
N.~Arkani-Hamed and J.~Maldacena,
``Cosmological Collider Physics,''
[\href{https://arxiv.org/abs/1503.08043}{{\tt arXiv:1503.08043 [hep-th]}}].

\bibitem{Arkani-Hamed:2018kmz}
N.~Arkani-Hamed, D.~Baumann, H.~Lee and G.~L.~Pimentel,
``The Cosmological Bootstrap: Inflationary Correlators from Symmetries and Singularities,''
JHEP \textbf{04} (2020), 105
[\href{https://arxiv.org/abs/1811.00024}{{\tt arXiv:1811.00024  [hep-th]}}].

\bibitem{Sleight:2019mgd}
C.~Sleight,
``A Mellin Space Approach to Cosmological Correlators,''
JHEP \textbf{01} (2020), 090
[\href{https://arxiv.org/abs/1906.12302}{{\tt arXiv:1906.12302 [hep-th]}}].

\bibitem{Sleight:2019hfp}
C.~Sleight and M.~Taronna,
``Bootstrapping Inflationary Correlators in Mellin Space,''
JHEP \textbf{02} (2020), 098
[\href{https://arxiv.org/abs/1907.01143}{{\tt arXiv:1907.01143 [hep-th]}}].

\bibitem{Baumann:2019oyu}
D.~Baumann, C.~Duaso Pueyo, A.~Joyce, H.~Lee and G.~L.~Pimentel,
``The Cosmological Bootstrap: Weight-Shifting Operators and Scalar Seeds,''
[\href{https://arxiv.org/abs/1910.14051}{{\tt arXiv:1910.14051 [hep-th]}}].

\bibitem{Baumann:2020dch}
D.~Baumann, C.~Duaso Pueyo, A.~Joyce, H.~Lee and G.~L.~Pimentel,
``The Cosmological Bootstrap: Spinning Correlators from Symmetries and Factorization,''
[\href{https://arxiv.org/abs/2005.04234}{{\tt arXiv:2005.04234 [hep-th]}}].









\bibitem{Huh:2013vga}
Y.~Huh, P.~Strack and S.~Sachdev,
``Conserved current correlators of conformal field theories in 2+1 dimensions,''
Phys. Rev. B \textbf{88}, 155109 (2013)
[erratum: Phys. Rev. B \textbf{90}, no.19, 199902 (2014)]
[\href{https://arxiv.org/abs/1307.6863}{{\tt arXiv:1307.6863 [cond-mat.str-el]}}].



\bibitem{Chowdhury:2012km}
D.~Chowdhury, S.~Raju, S.~Sachdev, A.~Singh and P.~Strack,
``Multipoint correlators of conformal field theories: implications for quantum critical transport,''
Phys. Rev. B \textbf{87}, no.8, 085138 (2013)
[\href{https://arxiv.org/abs/1210.5247}{{\tt arXiv:1210.5247 [cond-mat.str-el]}}].


\bibitem{Gillioz:2016jnn}
M.~Gillioz, X.~Lu and M.~A.~Luty,
``Scale Anomalies, States, and Rates in Conformal Field Theory,''
JHEP \textbf{04}, 171 (2017)
[\href{https://arxiv.org/abs/1612.07800}{{\tt arXiv:1612.07800 [hep-th]}}].



\bibitem{Coriano:2017mux}
C.~Coriano, M.~M.~Maglio and E.~Mottola,
``TTT in CFT: Trace Identities and the Conformal Anomaly Effective Action,''
Nucl. Phys. B \textbf{942} (2019), 303-328
[\href{https://arxiv.org/abs/1703.08860}{{\tt arXiv:1703.08860 [hep-th]}}].






\bibitem{Gillioz:2018kwh}
M.~Gillioz, X.~Lu and M.~A.~Luty,
``Graviton Scattering and a Sum Rule for the c Anomaly in 4D CFT,''
JHEP \textbf{09}, 025 (2018)
[\href{https://arxiv.org/abs/1801.05807}{{\tt arXiv:1801.05807 [hep-th]}}].


\bibitem{Coriano:2018zdo}
C.~Corian\`o and M.~M.~Maglio,
``Renormalization, Conformal Ward Identities and the Origin of a Conformal Anomaly Pole,''
Phys. Lett. B \textbf{781}, 283-289 (2018)
[\href{https://arxiv.org/abs/1802.01501}{{\tt arXiv:1802.01501 [hep-th]}}].


\bibitem{Coriano:2020ees}
C.~Corianò and M.~M.~Maglio,
``Conformal Field Theory in Momentum Space and Anomaly Actions in Gravity: The Analysis of 3- and 4-Point Functions,''
[\href{https://arxiv.org/abs/2005.06873}{{\tt arXiv:2005.06873 [hep-th]}}].



\bibitem{Anand:2019lkt}
N.~Anand, Z.~U.~Khandker and M.~T.~Walters,
``Momentum space CFT correlators for Hamiltonian truncation,''
JHEP \textbf{10}, 095 (2020)
[\href{https://arxiv.org/abs/1911.02573}{{\tt arXiv:1911.02573 [hep-th]}}].


\bibitem{Katz:2016hxp}
E.~Katz, Z.~U.~Khandker and M.~T.~Walters,
``A Conformal Truncation Framework for Infinite-Volume Dynamics,''
JHEP \textbf{07}, 140 (2016)
[\href{https://arxiv.org/abs/1604.01766}{{\tt arXiv:1604.01766 [hep-th]}}].



\bibitem{Polyakov:1974gs}
A.~M.~Polyakov,
``Non-Hamiltonian approach to conformal quantum field theory,''
Zh. Eksp. Teor. Fiz. \textbf{66}, 23-42 (1974)


\bibitem{Raju:2012zr}
S.~Raju,
``New Recursion Relations and a Flat Space Limit for AdS/CFT Correlators,''
Phys. Rev. D \textbf{85}, 126009 (2012)
[\href{https://arxiv.org/abs/1201.6449}{{\tt arXiv:1201.6449 [hep-th]}}].



\bibitem{Penedones:2010ue}
J.~Penedones,
``Writing CFT correlation functions as AdS scattering amplitudes,''
JHEP \textbf{03}, 025 (2011)
[\href{https://arxiv.org/abs/1011.1485}{{\tt arXiv:1011.1485 [hep-th]}}].



\bibitem{Fitzpatrick:2011hu}
A.~L.~Fitzpatrick and J.~Kaplan,
``Analyticity and the Holographic S-Matrix,''
JHEP \textbf{10}, 127 (2012)
[\href{https://arxiv.org/abs/1111.6972}{{\tt arXiv:1111.6972 [hep-th]}}].



\bibitem{Gary:2009ae}
M.~Gary, S.~B.~Giddings and J.~Penedones,
``Local bulk S-matrix elements and CFT singularities,''
Phys. Rev. D \textbf{80}, 085005 (2009)
[\href{https://arxiv.org/abs/0903.4437}{{\tt arXiv:0903.4437 [hep-th]}}].



\bibitem{Gary:2009mi}
M.~Gary and S.~B.~Giddings,
``The Flat space S-matrix from the AdS/CFT correspondence?,''
Phys. Rev. D \textbf{80}, 046008 (2009)
[\href{https://arxiv.org/abs/0904.3544}{{\tt arXiv:0904.3544 [hep-th]}}].


\bibitem{Komatsu:2020sag}
S.~Komatsu, M.~F.~Paulos, B.~C.~Van Rees and X.~Zhao,
``Landau diagrams in AdS and S-matrices from conformal correlators,''
JHEP \textbf{11}, 046 (2020)
[\href{https://arxiv.org/abs/2007.13745}{{\tt arXiv:2007.13745 [hep-th]}}].








\bibitem{Aharony:2011jz}
O.~Aharony, G.~Gur-Ari and R.~Yacoby,
``$d=3$ Bosonic Vector Models Coupled to Chern-Simons Gauge Theories,''
JHEP \textbf{03} (2012), 037
[\href{https://arxiv.org/abs/1110.4382}{{\tt arXiv:1110.4382 [hep-th]}}].



\bibitem{Giombi:2011kc}
S.~Giombi, S.~Minwalla, S.~Prakash, S.~P.~Trivedi, S.~R.~Wadia and X.~Yin,
``Chern-Simons Theory with Vector Fermion Matter,''
Eur. Phys. J. C \textbf{72} (2012), 2112
[\href{https://arxiv.org/abs/1110.4386}{{\tt arXiv:1110.4386 [hep-th]}}].

\bibitem{Maldacena:2012sf}
J.~Maldacena and A.~Zhiboedov,
``Constraining conformal field theories with a slightly broken higher spin symmetry,''
Class. Quant. Grav. \textbf{30} (2013), 104003
[\href{https://arxiv.org/abs/1204.3882}{{\tt arXiv:1204.3882 [hep-th]}}].




\bibitem{Aharony:2012nh}
O.~Aharony, G.~Gur-Ari and R.~Yacoby,
``Correlation Functions of Large N Chern-Simons-Matter Theories and Bosonization in Three Dimensions,''
JHEP \textbf{12} (2012), 028
[\href{https://arxiv.org/abs/1207.4593}{{\tt arXiv:1207.4593 [hep-th]}}].

\bibitem{GurAri:2012is}
G.~Gur-Ari and R.~Yacoby,
``Correlators of Large N Fermionic Chern-Simons Vector Models,''
JHEP \textbf{02} (2013), 150
[\href{https://arxiv.org/abs/1211.1866}{{\tt arXiv:1211.1866 [hep-th]}}].



\bibitem{Giombi:2016zwa}
S.~Giombi, V.~Gurucharan, V.~Kirilin, S.~Prakash and E.~Skvortsov,
``On the Higher-Spin Spectrum in Large N Chern-Simons Vector Models,''
JHEP \textbf{01} (2017), 058
[\href{https://arxiv.org/abs/1610.08472}{{\tt arXiv:1610.08472 [hep-th]}}].

\bibitem{Aharony:2018pjn}
O.~Aharony, S.~Jain and S.~Minwalla,
``Flows, Fixed Points and Duality in Chern-Simons-matter theories,''
JHEP \textbf{12} (2018), 058
[\href{https://arxiv.org/abs/1808.03317}{{\tt arXiv:1808.03317 [hep-th]}}].


\bibitem{Kalloor:2019xjb}
R.~R.~Kalloor,
``Four-point functions in large $N$ Chern-Simons fermionic theories,''
JHEP \textbf{10} (2020), 028
[\href{https://arxiv.org/abs/1910.14617}{{\tt arXiv:1910.14617 [hep-th]}}].


%
%

\bibitem{Osborn:1993cr}
H.~Osborn and A.~C.~Petkou,
``Implications of conformal invariance in field theories for general dimensions,''
Annals Phys. \textbf{231} (1994), 311-362
[\href{https://arxiv.org/abs/hep-th/9307010}{{\tt arXiv:9307010 [hep-th]}}].


\bibitem{Soda:2011am}
J.~Soda, H.~Kodama and M.~Nozawa,
``Parity Violation in Graviton Non-gaussianity,''
JHEP \textbf{08}, 067 (2011)
[\href{https://arxiv.org/abs/1106.3228}{{\tt arXiv:1106.3228 [hep-th]}}].

%
%


\bibitem{Lue:1998mq}
A.~Lue, L.~M.~Wang and M.~Kamionkowski,
``Cosmological signature of new parity violating interactions,''
Phys. Rev. Lett. \textbf{83}, 1506-1509 (1999)
[\href{https://arxiv.org/abs/astro-ph/9812088}{{\tt arXiv:9812088 [astro-ph]}}].



\bibitem{Shiraishi:2014roa}
M.~Shiraishi, M.~Liguori and J.~R.~Fergusson,
``General parity-odd CMB bispectrum estimation,''
JCAP \textbf{05}, 008 (2014)
[\href{https://arxiv.org/abs/1403.4222}{{\tt arXiv:1403.4222 [astro-ph.CO]}}].



\bibitem{Shiraishi:2014ila}
M.~Shiraishi, M.~Liguori and J.~R.~Fergusson,
``Observed parity-odd CMB temperature bispectrum,''
JCAP \textbf{01}, 007 (2015)
[\href{https://arxiv.org/abs/1409.0265}{{\tt arXiv:1409.0265 [astro-ph.CO]}}].



\bibitem{Costa:2011mg}
M.~S.~Costa, J.~Penedones, D.~Poland and S.~Rychkov,
``Spinning Conformal Correlators,''
JHEP \textbf{11} (2011), 071
[\href{https://arxiv.org/abs/1107.3554}{{\tt arXiv:1107.3554 [hep-th]}}].



\bibitem{Costa:2011dw}
M.~S.~Costa, J.~Penedones, D.~Poland and S.~Rychkov,
``Spinning Conformal Blocks,''
JHEP \textbf{11} (2011), 154
[\href{https://arxiv.org/abs/1109.6321}{{\tt arXiv:1109.6321 [hep-th]}}].


\bibitem{Karateev:2017jgd}
D.~Karateev, P.~Kravchuk and D.~Simmons-Duffin,
``Weight Shifting Operators and Conformal Blocks,''
JHEP \textbf{02} (2018), 081
[\href{https://arxiv.org/abs/1706.07813}{{\tt arXiv:1706.07813 [hep-th]}}].


\bibitem{Bzowski:2020lip}
A.~Bzowski,
``TripleK: A Mathematica package for evaluating triple-$K$ integrals and conformal correlation functions,''
Comput. Phys. Commun. \textbf{258} (2021), 107538
[\href{https://arxiv.org/abs/2005.10841}{{\tt arXiv:2005.10841 [hep-th]}}].







%


%



%
%
%



%


\bibitem{Giombi:2011rz}
S.~Giombi, S.~Prakash and X.~Yin,
``A Note on CFT Correlators in Three Dimensions,''
JHEP \textbf{07} (2013), 105
[\href{https://arxiv.org/abs/2005.10841}{{\tt arXiv:1104.4317 [hep-th]}}].


\bibitem{todorov}
V.~K.~Dobrev, V.~B.~Petkova, S.~G.~Petrova and I.~T.~Todorov, 
``Dynamical Derivation Of Vacuum Operator Product Expansion In Euclidean Conformal Quantum Field Theory," 
Phys. Rev. D 13, 887 (1976).

\bibitem{wip}
work to appear soon.
%

\bibitem{Aharony:2019mbc}
O.~Aharony and A.~Sharon,
``Large N renormalization group flows in 3d $ \mathcal{N} $ = 1 Chern-Simons-Matter theories,''
JHEP \textbf{07} (2019), 160
[\href{https://arxiv.org/abs/1905.07146}{{\tt arXiv:1905.07146 [hep-th]}}].


\bibitem{Inbasekar:2019wdw}
K.~Inbasekar, S.~Jain, V.~Malvimat, A.~Mehta, P.~Nayak and T.~Sharma,
``Correlation functions in ${\cal N}=2$ Supersymmetric vector matter Chern-Simons theory,''
JHEP \textbf{04} (2020), 207
[\href{https://arxiv.org/abs/1907.11722}{{\tt arXiv:1907.11722 [hep-th]}}].





%
\end{thebibliography}

\end{document}